\documentclass[journal,draftclsnofoot,onecolumn,12pt]{IEEEtran}
\IEEEoverridecommandlockouts
\usepackage{cite}
\usepackage{graphicx}
\usepackage{amsmath}
\usepackage{amssymb}
\usepackage{algorithmicx}
\usepackage{algorithm}
\usepackage{color}
\usepackage{pdfcomment}
\usepackage{bookmark}
\usepackage{enumerate}

\newcommand{\tabincell}[2]{\begin{tabular}{@{}#1@{}}#2\end{tabular}}  

\allowdisplaybreaks

\begin{document}
\title{Path Loss Modeling and Measurements for Reconfigurable Intelligent Surfaces in the Millimeter-Wave Frequency Band}
\author{
\IEEEauthorblockN{Wankai Tang, Xiangyu Chen, Ming Zheng Chen, Jun Yan Dai, Yu Han,\\ Marco Di Renzo, Shi Jin, Qiang Cheng, and Tie Jun Cui}
\thanks{W. Tang, X. Chen, Y. Han, and S. Jin are with the National Mobile Communications Research Laboratory, Southeast University, Nanjing, China. M. Z. Chen, J. Y. Dai, Q. Cheng, and T. J. Cui are with the State Key Laboratory of Millimeter Waves, Southeast University, Nanjing, China. M. Di Renzo is with Universit\'e Paris-Saclay, CNRS, CentraleSup\'elec, Laboratoire des Signaux et Syst\`emes, 3 Rue Joliot-Curie, 91192 Gif-sur-Yvette, France.}
}
\maketitle
\vspace{-1cm}
\begin{abstract}
Reconfigurable intelligent surfaces (RISs) provide an interface between the electromagnetic world of wireless propagation environments and the digital world of information science. Simple yet sufficiently accurate path loss models for RISs are an important basis for theoretical analysis and optimization of RIS-assisted wireless communication systems. In this paper, we refine our previously proposed free-space path loss model for RISs to make it simpler, more applicable, and easier to use. The impact of the antenna's directivity of the transmitter, receiver, and the unit cells of the RIS on the path loss is explicitly formulated as an angle-dependent loss factor. The refined model gives more accurate estimates of the path loss of RISs comprised of unit cells with a deep sub-wavelength size. Based on the proposed model, the properties of a single unit cell are evaluated in terms of scattering performance, power consumption, and area, which allows us to unveil fundamental considerations for deploying RISs in high frequency bands. Two fabricated RISs operating in the millimeter-wave (mmWave) band are utilized to carry out a measurement campaign. The measurement results are shown to be in good agreement with the proposed path loss model.
In addition, the experimental results suggest an effective form to characterize the power radiation pattern of the unit cell for path loss modeling.
\end{abstract}
\begin{IEEEkeywords}
Path loss, reconfigurable intelligent surface, millimeter-wave, intelligent reflecting surface, metasurface, wireless propagation measurements.
\end{IEEEkeywords}

\section{Introduction}
Fifth-generation (5G) mobile communication networks are being deployed in many countries in the world. Base stations equipped with massive multiple-input multiple-output (MIMO) transceivers provide support for enhanced mobile broadband (eMBB) applications and several other emerging communication services. With the current adoption of 5G standards, researchers in academia and industry started turning their research interests towards the sixth-generation (6G) of mobile communication networks\cite{6G}.

The millimeter-wave (mmWave) and terahertz (THz) frequency bands promise to offer abundant spectrum and unprecedented peak data rates for supporting the services of 6G networks. However, a major issue for enabling wireless communications in the mmWave and THz bands is the large path loss, especially when the transmission path is blocked by obstacles\cite{combat}. Common solutions to alleviate this issue include increasing the transmit power and deploying additional network infrastructure, e.g., relays. The deployment of relays is, however, usually characterized by an increase of the network power consumption and an increase of the hardware complexity and cost. Therefore, it is necessary to explore new hardware architectures.

In recent years, reconfigurable intelligent surfaces (RISs) have emerged as a promising enabling technology for 6G networks\cite{RIS1,RIS2,RIS3,RIS4,RIS5}. An RIS is a thin surface that comprises a large number of sub-wavelength unit cells that have the capability of modifying the phase and amplitude response on the incident signals\cite{MetaCoding,MetaInfo,MetaNature}. The two-dimensional structure and reconfigurable electromagnetic response of the RIS facilitate its potential applications in future wireless networks. Recently, for example, RISs have been utilized to realize single-RF MIMO transmitters that offer an attractive architecture for designing spatial multiplexing MIMO systems at a low complexity, power consumption, and hardware cost\cite{Modu1,Modu2,ModuNSR,Modu3,Modu4}.

Endowed with advanced capabilities in terms of manipulating the electromagnetic waves and shaping the wireless channels, RISs can help controlling the otherwise uncontrollable wireless environment and are viewed as an enabler of the emerging paradigm known as ``smart radio environments''\cite{Channel1,Channel2,Channel3}. For example, RISs can perform analog beamforming by co-phasing the signals reflected by the unit cells, which enhances the received signals and helps combating the distance and blocking problems in wireless communications. Moreover, RISs introduce new degrees of freedom for shaping the wireless channels and are viewed as an enabler for several applications, which include wireless power transfer\cite{WPT}, environmental sensing and positioning\cite{Sense}, and secure communications\cite{Secure}. An important but open research issue in RIS-assisted wireless communications is to develop simple but sufficiently accurate path loss models, which are an essential pre-requisite for link budget analysis, for understanding the ultimate performance limits offered by RISs, and for optimizing the configuration and deployment of RISs.
\subsection{Related Works}\label{RelatedWork}
Due to the importance of characterizing the path loss of RIS-assisted communications, some researchers have recently conducted theoretical and experimental studies. In \cite{pathloss}, path loss models for RISs are derived, for different application scenarios, with the aid of channel measurements that are conducted in an anechoic chamber. The measurements unveil that the power scaling law of the reflected signal from an RIS depends on several key parameters
and the configuration of the RIS.
In particular, the measurement campaign in \cite{pathloss} is carried out by using three fabricated RISs that operate below 10.5 GHz. However, the proposed path loss model does not yield an explicit expression of the joint radiation pattern of the antennas of the transmitter/receiver and the unit cells of the RIS.
In \cite{Ellingson}, a physical model for the path loss of an RIS-assisted link is proposed by using antenna theory, which confirms the main findings in \cite{pathloss}.
In \cite{bridge}, the radiation density of the scattered field in the near-field and far-field of the RIS is calculated under the assumption of dipole antennas. The scaling laws of the path loss as a function of the transmission distances are discussed numerically but no explicit characterization of the unit cells of planar RISs is given.
In \cite{pathloss-spawc}, integral and approximated closed-form expressions of the path loss of a one-dimensional RIS in the far-field and near-field regions are given. The results are obtained by leveraging the general scalar theory of diffraction. Different scaling laws as a function of the transmission distances and the size of the RIS are observed in the near-field and far-field regions. In \cite{vector}, the analysis in \cite{pathloss-spawc} is generalized to two-dimensional RISs by using the vector generalization of Green's theorem. In particular, the scaling laws of the electric and magnetic fields as a function of the transmission distance and the size of the RIS are characterized. The approach is based on the theory of scattering, which implies that the directional radiation characteristics of the unit cells of the RIS are not explicitly taken into account, since a continuous impedance sheet model for the RIS is assumed. In \cite{NLOS}, passive reflectors are utilized to enhance the coverage of mmWave communications.
However, since passive reflectors are different from RISs, the analytical expressions in \cite{NLOS} cannot be directly used to characterize the path loss of RISs. In \cite{Coupling}, the authors introduce a path loss model that is based on the theory of mutual impedances. The end-to-end channel model is formulated in an algebraic form, which resembles a MIMO communication system. The approach considers the mutual coupling among RIS unit cells, however, it is not directly applicable to planar RISs. The path loss model in \cite{Coupling} has recently been employed in \cite{Coupling2} and \cite{Coupling3} for system optimization. A summary of research works on path loss modeling for RISs is available in Table I of \cite{Modelsummary}.
\subsection{Main Contributions}\label{MainContributions}
Motivated by the limitations in existing works,
in this paper, we refine the free-space path loss model previously proposed in \cite{pathloss} for RIS-assisted wireless networks. New measurement results conducted in the mmWave band, which are reported for the first time in this paper, and the measurement results reported before corroborate the accuracy of the refined path loss model. The main contributions of this paper are summarized as follows:
\begin{enumerate}
\item We introduce model refinements to our previously proposed free-space path loss model for RISs in order to make it simpler and easier to use. The impact of the radiation patterns of the transmit/receive antennas and the unit cells of the RIS is explicitly formulated. The accuracy of the proposed path loss model for application to RISs comprised of unit cells with a deep sub-wavelength size is improved. 

\item We evaluate the properties of a single unit cell in terms of scattering performance, power, and area, as it is the basic element of an RIS. It is shown that the energy efficiency of a unit cell is inversely proportional to the square of the operating frequency, while the power consumption per unit area is proportional to the square of the operating frequency. These results reveal fundamental challenges for deploying RISs in high frequency bands.

\item We report the world's first measurement campaign in the mmWave frequency band to validate the path loss model for RIS-assisted wireless communications. It is experimentally proved that the relation between the scattering gain of a unit cell and its size can be properly modeled by using a metal plate as benchmark. The measurements also suggest that $\cos\theta$ is a practical and effective expression to model the shape of the power radiation pattern of a sub-wavelength unit cell. In general, the experimental results are in good agreement with the proposed analytical formulation of the path loss of an RIS-assisted link.


\end{enumerate}
\section{Brief Review of Path Loss Models for RISs}
In this section, we give a brief summary of the free-space path loss models for RISs proposed in \cite{pathloss}, and discuss their limitations and how they can be improved.

\subsection{System Description}\label{SystemDescription}
\begin{figure}
\centering
\includegraphics[scale = 0.78]{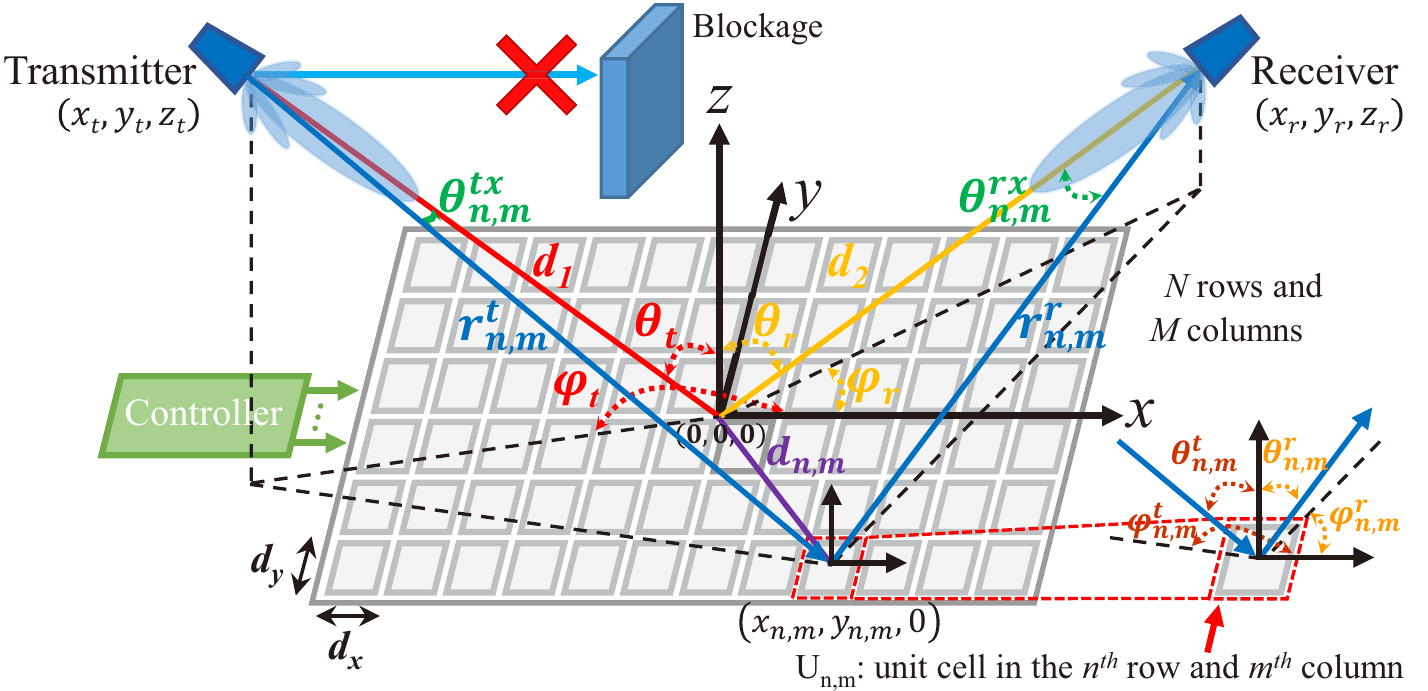}
\caption{The considered RIS-assisted wireless communication system.}
\label{systemdescription}
\end{figure}
In this paper, we consider the general RIS-assisted wireless communication system illustrated in Fig. \ref{systemdescription}, which can be applied in the sub-6 GHz and mmWave frequency bands. The direct link between the transmitter and the receiver is assumed to be completely blocked by obstacles, since it has already been well studied in the literature\cite{Book}. The RIS is deployed in the x-y plane, and the center of the RIS is placed at $(0, 0, 0)$, i.e., the origin of the  Cartesian coordinate system. The RIS is comprised of unit cells that are evenly distributed in $N$ rows and $M$ columns. The width and length of each unit cell is denoted by $d_x$ and $d_y$, respectively. $\rm{U}_{n,m}$ denotes the unit cell in the $n^{th}$ row and $m^{th}$ column, where $n \in \left[ {1,N} \right]$ and $m \in \left[ {1,M} \right]$. Each unit cell is usually made of a scattering element (e.g., a patch) and tunable components (e.g., a positive intrinsic negative diode). By appropriately tuning the tunable components, the reflection coefficient ${\varGamma _{n,m}}$ of the unit cell $\rm{U}_{n,m}$ can be flexibly adjusted. As shown in Fig. \ref{systemdescription}, $d_{1}$, $d_{2}$, $\theta_{t}$, $\varphi_{t}$, $\theta_{r}$, and $\varphi_{r}$ denote the distance between the transmitter and the center of the RIS, the distance between the receiver and the center of the RIS, the elevation angle and the azimuth angle from the center of the RIS to the transmitter, the elevation angle and the azimuth angle from the center of the RIS to the receiver, respectively. As far as the geometric description of each unit cell is concerned, we use the notation $r_{n,m}^t$, $r_{n,m}^r$, $\theta_{n,m}^t$, $\varphi_{n,m}^t$, $\theta_{n,m}^r$, and $\varphi_{n,m}^r$ to denote the distance between the transmitter and $\rm{U}_{n,m}$, the distance between the receiver and $\rm{U}_{n,m}$, the elevation angle and the azimuth angle from $\rm{U}_{n,m}$ to the transmitter, the elevation angle and the azimuth angle from $\rm{U}_{n,m}$ to the receiver, respectively.

As illustrated in Fig. \ref{systemdescription}, the transmitter located at $(x_t, y_t, z_t)$ transmits a signal to the receiver located at $(x_r, y_r, z_r)$ through the RIS-assisted link. The transmitted signal with power $P_t$ and wavelength $\lambda$ is reflected by the RIS before reaching the receiver. $F({\theta ,\varphi})$, $G$, ${F^{tx}}(\theta,\varphi)$, $G_t$, ${F^{rx}}(\theta,\varphi)$, and $G_r$ denote the power radiation pattern and the scattering gain of the unit cell, power radiation pattern and the gain of the transmit antenna, the power radiation pattern and the gain of the receive antenna, respectively, where $\theta$ and $\varphi$ are the elevation and azimuth angles from the unit cell/antenna to a certain direction of transmission/reception. $\theta_{n,m}^{tx}$, $\varphi_{n,m}^{tx}$, $\theta_{n,m}^{rx}$, and $\varphi_{n,m}^{rx}$ denote the elevation angle and the azimuth angle from the transmit antenna to the unit cell $\rm{U}_{n,m}$, and the elevation angle and the azimuth angle from the receive antenna to the unit cell $\rm{U}_{n,m}$, respectively. Notation and definitions are summarized in Table \ref{Notations}.

\begin{table}
\centering
\footnotesize
\caption{Notation and definitions.}\label{Notations}
\begin{tabular}{|c|l|c|l|}
\hline
\textbf{Notation} & \textbf{Definition}  & \textbf{Notation} & \textbf{Definition}\\
\hline
$N$ &  Number of rows of unit cells & $M$ &  Number of columns of unit cells\\
\hline
$\rm{U}_{n,m}$ &  Unit cell in the $n^{th}$ row and $m^{th}$ column & $\varGamma _{n,m}$ & Reflection coefficient of $\rm{U}_{n,m}$\\
\hline
$d_x$  & Width of each unit cell & $d_y$ & Length of each unit cell\\
\hline
$(x_{n,m}, y_{n,m}, 0)$ & Position of the unit cell $\rm{U}_{n,m}$ & ${d_{n,m}}$ & \tabincell{l}{Distance between $\rm{U}_{n,m}$ and the center \\of the RIS. ${d_{n,m}} = \sqrt {x_{n,m}^2 + y_{n,m}^2}$}\\
\hline
$(x_t, y_t, z_t)$ & Position of the transmitter & $(x_r, y_r, z_r)$ & Position of the receiver\\
\hline
$d_{1}$ & \tabincell{l}{Distance between the transmitter and the \\center of the RIS. ${d_1} = \sqrt {x_t^2 + y_t^2 + z_t^2}$} & $d_{2}$ & \tabincell{l}{Distance between the receiver and the \\center of the RIS. ${d_2} = \sqrt {x_r^2 + y_r^2 + z_r^2}$}\\
\hline
$(\theta_{t}, \varphi_{t})$ & \tabincell{l}{Elevation angle and azimuth angle from \\the center of the RIS to the transmitter} & $r_{n,m}^t$ & \tabincell{l}{Distance between the transmitter and $\rm{U}_{n,m}$ \\ $r_{n,m}^t{=}\sqrt {{{\left( {{x_t}{-}{x_{n,m}}} \right)}^2} + {{\left( {{y_t}{-}{y_{n,m}}} \right)}^2} + z_t^2}$}\\
\hline
$(\theta_{r}, \varphi_{r})$ & \tabincell{l}{Elevation angle and azimuth angle from \\the center of the RIS to the receiver} & $r_{n,m}^r$ & \tabincell{l}{Distance between the receiver and $\rm{U}_{n,m}$\\ $r_{n,m}^r{=}\sqrt {{{\left( {{x_r}{-}{x_{n,m}}} \right)}^2} + {{\left( {{y_r}{-}{y_{n,m}}} \right)}^2} + z_r^2}$}\\
\hline
$(\theta_{n,m}^t, \varphi_{n,m}^t)$ & \tabincell{l}{Elevation angle and azimuth angle from \\$\rm{U}_{n,m}$ to the transmitter} & $F({\theta ,\varphi })$ & \tabincell{l}{Normalized power radiation pattern of each \\unit cell}\\
\hline
$(\theta_{n,m}^r, \varphi_{n,m}^r)$& \tabincell{l}{Elevation angle and azimuth angle from\\ $\rm{U}_{n,m}$ to the receiver} & $G$ &  Scattering gain of a single unit cell \\
\hline
${F^{tx}}(\theta,\varphi)$ & \tabincell{l}{Normalized power radiation pattern of \\the transmit antenna} & ${F^{rx}}(\theta,\varphi)$ & \tabincell{l}{Normalized power radiation pattern of the \\receive antenna}\\
\hline
$G_t$ & Gain of the transmit antenna & $G_r$ & Gain of the receive antenna\\
\hline
$(\theta_{n,m}^{tx}, \varphi_{n,m}^{tx})$& \tabincell{l}{Elevation angle and azimuth angle from \\the transmit antenna to $\rm{U}_{n,m}$} & $\lambda$ & Wavelength of the transmitted signal\\
\hline
$(\theta_{n,m}^{rx}, \varphi_{n,m}^{rx})$& \tabincell{l}{Elevation angle and azimuth angle from \\the receive antenna to $\rm{U}_{n,m}$}& $P_t$ & Power of the transmitted signal \\
\hline
\end{tabular}
\end{table}
\subsection{Previously Proposed Path Loss Models for RIS-Assisted Transmission}\label{PreviousModel}
In our prior work \cite{pathloss}, a general free-space path loss model that accounts for the physics and electromagnetic nature of the RIS was derived. Based on the general model, free-space path loss models for three different specific scenarios were obtained.
In this section, we briefly review these free-space path loss models and discuss how they can be further improved.

\subsubsection{General Case}\label{Case0}
According to Theorem 1 in \cite{pathloss}, the received signal power $P_r$ of an RIS-assisted link, as a function of several system parameters, can be formulated as follows
\begin{equation}\label{s1}
{P_r}{=}{P_t}\frac{{G_t}{G_r}G{d_x}{d_y}{\lambda ^2}}{64{\pi ^3}}
{\left| {\sum\limits_{m = 1}^{M} {\sum\limits_{n = 1}^{N} {\frac{\sqrt {F_{n,m}^{combine}}\  {\varGamma _{n,m}}}{r_{n,m}^tr_{n,m}^r}{e^{\frac{- j2\pi (r_{n,m}^t + r_{n,m}^r)}{\lambda }}}}}} \right|^2},
\end{equation}
where ${F_{n,m}^{combine}}{=}{F^{tx}}\left(\theta _{n,m}^{tx},\varphi _{n,m}^{tx}\right)F\left(\theta _{n,m}^t,\varphi _{n,m}^t\right)F\left(\theta _{n,m}^r, \varphi _{n,m}^r\right){F^{rx}}\left(\theta _{n,m}^{rx},\varphi _{n,m}^{rx}\right)$ accounts for the effect of the normalized power radiation patterns (i.e., the directivities) of the transmit antenna, the unit cell, and the receive antenna.
The path loss corresponding to (\ref{s1}) is
\begin{equation}\label{s2}
PL_{general} = \frac{P_t}{P_r} = \frac{64{\pi ^3}}{{G_t}{G_r}G{d_x}{d_y}{\lambda ^2}{\left| {\sum\limits_{m = 1}^{M} {\sum\limits_{n = 1}^{N} {\frac{\sqrt {F_{n,m}^{combine}}\  {\varGamma _{n,m}}}{r_{n,m}^tr_{n,m}^r}{e^{\frac{- j2\pi (r_{n,m}^t + r_{n,m}^r)}{\lambda }}}}}} \right|^2}},
\end{equation}
which is referred to as the general free-space path loss model for RIS-assisted communications. Equation (\ref{s2}) can be applied to arbitrarily configured RISs in the near-field and far-field regions.
Based on (\ref{s2}), path loss models for three typical scenarios were proposed in\cite{pathloss}.
\subsubsection{RIS-Assisted Far-Field Beamforming}\label{Case1}
In the far-field region, the RIS can beamform the incident signals towards a specific direction, so as to enhance the received signal power. In \cite{pathloss}, the free-space path loss for RIS-assisted far-field beamforming is formulated as follows
\begin{equation}\label{s3}
PL_{farfield}^{beam}=\frac{P_t}{P_r}=\frac{{64{\pi ^3}{{({d_1}{d_2})}^2}}}{{{G_t}{G_r}G{M^2}{N^2}{d_x}{d_y}{\lambda ^2}F({\theta _t},{\varphi _t})F({\theta _r},{\varphi _r}){A^2}}},
\end{equation}
where $A$ is the amplitude of the reflection coefficient of each unit cell, which is assumed to be identical for all the unit cells of the RIS.

\subsubsection{RIS-Assisted Near-Field Focusing}\label{Case2}
In the near-field region, the signals reflected by the unit cells of the RIS can be focused to maximize the received signal power towards a specific user. In this case, in particular, the RIS needs to be optimized by considering that the plane wave approximation is not applicable anymore. Therefore, the different transmission distances and different angles of incidence and reflection between the unit cells of the RIS and the transmit and receive antennas need to be taken into account. In \cite{pathloss}, the free-space path loss for RIS-assisted near-field focusing is formulated as follows
\begin{equation}\label{s4}
PL_{nearfield}^{focus}=\frac{P_t}{P_r}=\frac{{64{\pi ^3}}}{{{G_t}{G_r}G{d_x}{d_y}{\lambda ^2}{A^2}{{\left| {\sum\limits_{m = 1}^{M} {\sum\limits_{n = 1}^{N} { \frac{\sqrt {{F_{n,m}^{combine}}}}{{r_{n,m}^tr_{n,m}^r}}} } } \right|}^2}}}.
\end{equation}

\subsubsection{RIS-Assisted Near-Field Broadcasting}\label{Case3}
In the near-field region, the reflected signals can be broadcasted by the RIS to provide coverage to several users in a specific area. In \cite{pathloss}, the free-space path loss for RIS-assisted near-field broadcasting is formulated as follows
\begin{equation}\label{s5}
PL_{nearfield}^{broadcast} \approx \frac{{16{\pi ^2}{{({d_1} + {d_2})}^2}}}{{{G_t}{G_r}{\lambda ^2}{A^2}}}.
\end{equation}

As for the design of the phase shifts $\angle {\varGamma_{n,m}}$ in (\ref{s3})-(\ref{s5}), the readers are referred to \cite{pathloss} for further details. Equations (\ref{s3})-(\ref{s5}) reveal that RIS-assisted links have different overall path loss functions that depend on the configuration of the unit cells and the propagation region. In particular, the overall path loss is proportional to $({d_1}{d_2})^2$ in the RIS-assisted far-field beamforming scenario, and is proportional to $(d_{1}+ d_{2})^2$ in the RIS-assisted near-field broadcasting scenario.
\subsection{Limitations of Existing Path Loss Models}\label{WeakAspects}
In this paper, we show that the path loss model in \cite{pathloss} can be further improved in order to make it more accurate and easier to use. In particular, we focus on two major aspects.
\subsubsection{Joint Radiation Pattern of Antennas and Unit Cells}\label{ComplexPattern}
The existing works mentioned in Section \ref{RelatedWork} on modeling the path loss of RIS-assisted links often assume that the transmit and receive antennas are isotropic, which results in ignoring the impact of the antennas' directivity. Although the authors of \cite{pathloss} have introduced the factor $F_{n,m}^{combine}$ that accounts for the joint normalized power radiation pattern of the transmit antenna, the unit cell, and the receive antenna, two major limitations still exist: (i) the impact of the geometry of the RIS-assisted link on $F_{m,n}^{combine}$ is not explicitly formulated; and (ii) an explicit expression for the power radiation pattern $F(\theta, \varphi)$ of the unit cell is neither explicitly discussed nor validated through measurements.

\subsubsection{Relation Between the Scattering Gain of the Unit Cell and Its Size}\label{InfiniteGain}
In \cite{pathloss}, no explicit relation between the scattering gain $G$ and the size $d_x d_y$ of the unit cell is given. This may lead to a misinterpretation of the scaling laws of the path loss as a function of the size and the number of unit cells. For example, the path loss model in (\ref{s3}) depends on the factor $G M N S_{RIS}$, where $S_{RIS} = M N d_x d_y$ is the total area of the RIS. By keeping $S_{RIS}$ fixed, the received power may increase without bound if the number of unit cells $M N$ increases (i.e., $d_x d_y$ decreases). This apparent inconsistency originates from the fact that the scattering gain $G$ and the size $d_x d_y$ of the unit cell are not independent of each other. Some related works like \cite{Ellingson} indicate that the metal plate scattering model may serve as a useful benchmark to identify the relation between $G$ and $d_x d_y$. However, further analysis and experimental validation are needed.

\section{Refined Path Loss Models for RISs}\label{SectionIII}
In this section, the two aforementioned limitations are addressed individually and refined path loss models for RISs are introduced. Several insights from the models are discussed as well.
\subsection{Explicit Formulation of the Joint Radiation Pattern}\label{ModelRefinement1}
According to (\ref{s1}), the joint normalized power radiation pattern is defined as follows
\begin{equation}\label{s6}
{F_{n,m}^{combine}}{=}{F^{tx}}\left(\theta _{n,m}^{tx},\varphi _{n,m}^{tx}\right)F\left(\theta _{n,m}^t,\varphi _{n,m}^t\right)F\left(\theta _{n,m}^r, \varphi _{n,m}^r\right){F^{rx}}\left(\theta _{n,m}^{rx},\varphi _{n,m}^{rx}\right),
\end{equation}
where ${F^{tx}}\left(\theta,\varphi\right)$, ${F^{rx}}\left(\theta,\varphi\right)$, and ${F}\left(\theta,\varphi\right)$ are the normalized power radiation patterns of the transmit antenna, the receive antenna, and the unit cell, respectively. The following proposition provides an explicit expression for $F_{n,m}^{combine}$ that accounts for the antenna gains (i.e., $G_t$ and $G_r$) and the geometry of the RIS-assisted communication link.

{\emph{Proposition 1:}} Assume that the directions of peak radiation of both the transmit and receive antennas point towards the center of the RIS, the joint normalized power radiation pattern in (\ref{s6}) can
be rewritten as
\begin{equation}\label{s23}
\begin{aligned}
&F_{n,m}^{combine}\mathop  = \limits^{\left( a \right)}{\left( {\cos \theta _{n,m}^{tx}} \right)^{^{(\frac{{{G_t}}}{2} - 1)}}}\left( {\cos \theta _{n,m}^t} \right)\left( {\cos \theta _{n,m}^r} \right){\left( {\cos \theta _{n,m}^{rx}} \right)^{^{(\frac{{{G_r}}}{2} - 1)}}} \\
&= {\left( {\frac{{{{\left( {{d_1}} \right)}^2} + {{\left( {r_{n,m}^t} \right)}^2} - {{\left( {{d_{n,m}}} \right)}^2}}}{{2{d_1}r_{n,m}^t}}} \right)^{(\frac{{{G_t}}}{2} - 1)}}\left( {\frac{{{z_t}}}{{r_{n,m}^t}}} \right)\left( {\frac{{{z_r}}}{{r_{n,m}^r}}} \right){\left( {\frac{{{{\left( {{d_2}} \right)}^2} + {{\left( {r_{n,m}^r} \right)}^2} - {{\left( {{d_{n,m}}} \right)}^2}}}{{2{d_2}r_{n,m}^r}}} \right)^{(\frac{{{G_r}}}{2} - 1)}},
\end{aligned}
\end{equation}
where (a) follows by using the normalized power radiation patterns of the transmit antenna, receive antenna, and unit cell in (\ref{s11}), (\ref{s12}), and (\ref{s20}), respectively. The distances $d_1$, $d_2$, $r_{n,m}^t$, $r_{n,m}^r$, and $d_{n,m}$, which is the distance from $\rm{U}_{n,m}$ to the center of the RIS, can be obtained based on the positions of the transmitter, the receiver, and the unit cell $\rm{U}_{n,m}$. For the convenience of the readers, the explicit expressions of these distances are summarized in Table \ref{Notations}.

\emph{Proof}: See Appendix A. \hfill $\blacksquare$

Proposition 1 explicitly formulates the joint normalized power radiation pattern as a function of ${\left( {\cos \theta _{n,m}^{tx}} \right)^{^{(\frac{{{G_t}}}{2} - 1)}}}$, ${\cos \theta _{n,m}^t}{\cos \theta _{n,m}^r}$, and ${\left( {\cos \theta _{n,m}^{rx}} \right)^{^{(\frac{{{G_r}}}{2} - 1)}}}$, which account for the impact of the directivity of the transmit antenna, the unit cell, and the receive antenna, respectively, on the path loss. The advantage of the analytical formulation in (\ref{s23}) is that all its constituent parameters can be easily calculated as summarized in Table \ref{Notations}. Proposition 1 unveils that $F_{n,m}^{combine}$ is an angle-dependent loss factor. In particular, (\ref{s23}) shows that the amount of power that the unit cell $\rm{U}_{n,m}$ is capable of sensing and reflecting depends on the location of $\rm{U}_{n,m}$ within the RIS with respect to the location of the transmitter and receiver. In particular, the amount of power sensed and reflected by the unit cells located closer to the edges of the RIS is smaller.
\subsection{Explicit Relation Between the Scattering Gain of the Unit Cell and Its Size }\label{ModelRefinement2}
In order to explicitly account for the relation between the scattering gain $G$ of a unit cell and its size $d_x d_y$, we benchmark the path loss of an RIS configured for beamforming in the far-field region with the path loss of a rectangular metal plate with the same size and shape as the RIS. The corresponding path loss is given in the following proposition and it is used next to refine the general path loss model introduced in \cite{pathloss}, i.e., (\ref{s2}) in this paper.

{\emph{Proposition 2:}} Assume that the amplitude of the reflection coefficient of all the unit cells of the RIS is identical and equal to $A$. The free-space path loss of an RIS-assisted communication link in the far-field beamforming scenario can be formulated as follows
\begin{equation}\label{s28}
\begin{aligned}
PL_{farfield-beam}^{refined}=\frac{{16{\pi ^2}{{({d_1}{d_2})}^2}}}{{{G_t}{G_r}{{\left( {MN{d_x}{d_y}} \right)}^2}F({\theta _t},{\varphi _t})F({\theta _r},{\varphi _r}){A^2}}}
\mathop  = \limits^{\left( a \right)} \frac{{16{\pi ^2}{{({d_1}{d_2})}^2}}}{{{G_t}{G_r}{{\left( {MN{d_x}{d_y}} \right)}^2}{\cos {\theta _t}}{\cos {\theta _r}}{A^2}}},
\end{aligned}
\end{equation}
where (a) follows by using the normalized power radiation pattern of the unit cell in (\ref{s20}).

\emph{Proof}: See Appendix B. \hfill $\blacksquare$

Proposition 2 reveals that the beamforming gain in the far-field beamforming scenario is determined by the square of the total geometric area of the RIS, i.e., ${{\left( {MN{d_x}{d_y}} \right)}^2}$.
Moreover, we observe that the path loss is related to both the angle of incidence $\theta_t$ and the angle of reception $\theta_r$, which ensures the proposed model fulfills the channel reciprocity condition. This is consistent with the findings reported in \cite{Reciprocity}. In particular, the path loss increases as $\theta_t$ and/or $\theta_r$ increase. Based on the proof in Appendix B, in addition, we find that the relation between the scattering gain and the size of the unit cell is $G = \frac{{4\pi {d_x}{d_y}}}{{{\lambda ^2}}}$.

Based on the benchmark given in Proposition 2, the following theorem yields a general expression of the path loss for RISs that accounts for the interplay between the scattering gain and the size of the unit cell.

{\emph{Theorem 1:}} By explicitly taking into account the relation between the scattering gain and the size of the unit cell, the general free-space path loss model for RIS-assisted wireless communication systems can be expressed as
\begin{equation}\label{s29}
PL_{general}^{refined} = \frac{P_t}{P_r} = \frac{16{\pi ^2}}{{G_t}{G_r}{\left({d_x}{d_y}\right)}^2{\left| {\sum\limits_{m = 1}^{M} {\sum\limits_{n = 1}^{N} {\frac{\sqrt {F_{n,m}^{combine}}\  {\varGamma _{n,m}}}{r_{n,m}^tr_{n,m}^r}{e^{\frac{- j2\pi (r_{n,m}^t + r_{n,m}^r)}{\lambda }}}}}} \right|^2}},
\end{equation}
where $F_{n,m}^{combine}$ is given in (\ref{s23}).

\emph{Proof}: It follows from (\ref{s2}) by using the identity $G = \frac{{4\pi {d_x}{d_y}}}{{{\lambda ^2}}}$, as dictated by Proposition 2.

It is worth noting that ${{d_x}{d_y}}$ is inversely proportional to the square of the operating frequency. Therefore, Theorem 1 reveals that the path loss of an RIS-assisted link is proportional to the fourth power of the operating frequency.

{\emph{Corollary 1:}} Assume that the amplitude of the reflection coefficient of all the unit cells is the same, i.e., $\left| {\varGamma _{n,m}} \right|=A$. Then, the refined free-space path loss model for RIS-assisted communications in the near-field focusing scenario can be expressed as
\begin{equation}\label{s30}
PL_{nearfield-focus}^{refined}=\frac{P_t}{P_r}=\frac{{16{\pi ^2}}}{{{G_t}{G_r}{\left({d_x}{d_y}\right)}^2{A^2}{{\left| {\sum\limits_{m = 1}^{M} {\sum\limits_{n = 1}^{N} { \frac{\sqrt {{F_{n,m}^{combine}}}}{{r_{n,m}^tr_{n,m}^r}}} } } \right|}^2}}}.
\end{equation}

\emph{Proof}: It follows from (\ref{s29}) by setting $\left| {{\Gamma _{n,m}}} \right| = A$ and $\angle {\varGamma _{n,m}} = {e^{\frac{{j2\pi (r_{n,m}^t + r_{n,m}^r)}}{\lambda }}}$.

Theorem 1, Corollary 1, and Proposition 2 give the proposed refined free-space path loss models of RIS-assisted communications for application to a general setup, to the near-field focusing setup, and to the far-field beamforming setup, respectively. The path loss model in (\ref{s5}) for the near-field broadcasting setup is obtained by using an asymptotic approximation based on geometric optics, which is not affected by the proposed modifications.

Based on the path loss model in Theorem 1, it is instructive to characterize the free-space path loss model of a single unit cell.

{\emph{Corollary 2:}} The free-space path loss model for single unit cell $\rm{U}_{n,m}$ can be formulated as
\begin{equation}\label{s31}
\begin{aligned}
PL_{U_{n,m}}&=\frac{P_t}{P_{n,m}^r}= \frac{{16{\pi ^2}}}{{{G_t}{G_r}{{\left( {{d_x}{d_y}} \right)}^2}{{\left| {\frac{{\sqrt {F_{n,m}^{combine}}\ {\varGamma_{n,m}}}}{{r_{n,m}^tr_{n,m}^r}}{e^{\frac{{ - j2\pi (r_{n,m}^t + r_{n,m}^r)}}{\lambda }}}} \right|}^2}}}\\
&= \frac{{16{\pi ^2}{{\left( {r_{n,m}^tr_{n,m}^r} \right)}^2}}}{{{G_t}{G_r}{{\left( {{d_x}{d_y}} \right)}^2}F_{n,m}^{combine}{{\left| {{\varGamma_{n,m}}} \right|}^2}}},
\end{aligned}
\end{equation}
where $F_{n,m}^{combine}$ is given in (\ref{s23}) and ${P_{n,m}^r}$ denotes the received power due to the reflection from $\rm{U}_{n,m}$.

\emph{Proof}: It follows by direct inspection of (\ref{s29}), by invoking the superposition principle.

Corollary 2 unveils that the path loss of the sub-channel provided by unit cell $\rm{U}_{n,m}$ is proportional to the square of the product of the distances from the transmitter/receiver to $\rm{U}_{n,m}$, and is inversely proportional to the square of the area of $\rm{U}_{n,m}$ and to the loss factor $F_{n,m}^{combine}$.
\subsection{Model Summary and Discussion}\label{ModelSummary}
\begin{table}
\centering
\footnotesize
\caption{Summary of the Refined Path Loss Models for RIS-Assisted Wireless Communications}\label{refinedmodelsummary}
\begin{tabular}{|c|c|c|c|}
\hline
\textbf{Model Name} & \tabincell{c}{\textbf{Model}\\\textbf{in \cite{pathloss}}}  & \tabincell{c}{\textbf{Refined}\\\textbf{Model}} & \tabincell{c}{\textbf{Notes}}\\
\hline
\tabincell{l}{General free-space path loss model for\\ RIS-assisted wireless communication}& \tabincell{l}{(\ref{s2})} & (\ref{s29}) & \tabincell{l}{It applies to both electrically small and large RISs.\\It applies in both the near-field and far-field regions of the RIS.\\$F_{n,m}^{combine}$ is given in (\ref{s23}) for the refined model.}\\
\hline
\tabincell{l}{General free-space path loss model for\\ a single unit cell of the RIS}& \tabincell{c}{Not\\explicitly\\ provided}& (\ref{s31}) & \tabincell{l}{It applies to both electrically small and large RISs.\\It applies in both the near-field and far-field regions of the RIS.\\$F_{n,m}^{combine}$ is given in (\ref{s23}) for the refined model.}\\
\hline
\tabincell{l}{Specific free-space path loss model for\\ RIS-assisted far-field beamforming}& \tabincell{l}{(\ref{s3})} & (\ref{s28}) & \tabincell{l}{It applies to both electrically small and large RISs.\\The transmitter and receiver are in the far-field region of the RIS.\\The design method for $\angle {\varGamma_{n,m}}$ is given by (9) in \cite{pathloss}.}\\
\hline
\tabincell{l}{Specific free-space path loss model for\\ RIS-assisted near-field focusing}& \tabincell{l}{(\ref{s4})} & (\ref{s30}) & \tabincell{l}{It applies to both electrically small and large RISs.\\The transmitter and the receiver are both or only one of them is\\ in the near-field region of the RIS.\\The design method for $\angle {\varGamma_{n,m}}$ is given by (12) in \cite{pathloss}.}\\
\hline
\tabincell{l}{Specific free-space path loss model for\\ RIS-assisted near-field broadcasting}& \tabincell{l}{(\ref{s5})} & (\ref{s5}) & \tabincell{l}{It applies only to electrically large RISs.\\The transmitter and the receiver are both or only one of them is\\ in the near-field region of the RIS.\\The design method for $\angle {\varGamma_{n,m}}$ is given by (16) in \cite{pathloss}.}\\
\hline
\end{tabular}
\end{table}
The proposed free-space path loss models for RIS-assisted wireless communications are summarized in Table \ref{refinedmodelsummary}. In particular, according to (\ref{s31}), the received signal that is due to the reflection from a single unit cell $\rm{U}_{n,m}$ of the RIS can be formulated as
\begin{equation}\label{s33}
s_{n,m}^r = {s_t}\frac{{\sqrt {{G_t}{G_r}} \overbrace {\left( {{d_x}{d_y}} \right)}^{{\text{Size}}}\overbrace {\sqrt {F_{n,m}^{combine}} }^{{\text{Angle-dependent factor}}}\overbrace {\left| {{\varGamma _{n,m}}} \right|}^{{\text{Efficiency}}}}}{{4\pi \underbrace {\left( {r_{n,m}^tr_{n,m}^r} \right)}_{{\text{Product distance}}}}}{e^{ - j\left( {\frac{{2\pi }}{\lambda }\left( {r_{n,m}^t + r_{n,m}^r} \right) - \angle {\varGamma _{n,m}}} \right)}},
\end{equation}
where $s_t$ is the transmitted signal. Based on (\ref{s33}), since a unit cell is the atomic element that constitutes an RIS, we can evaluate the fundamental tradeoffs offered by an RIS in terms of scattering performance, power consumption, and area (SPA) of the unit cell.

\begin{figure}
\centering
\includegraphics[scale = 0.475]{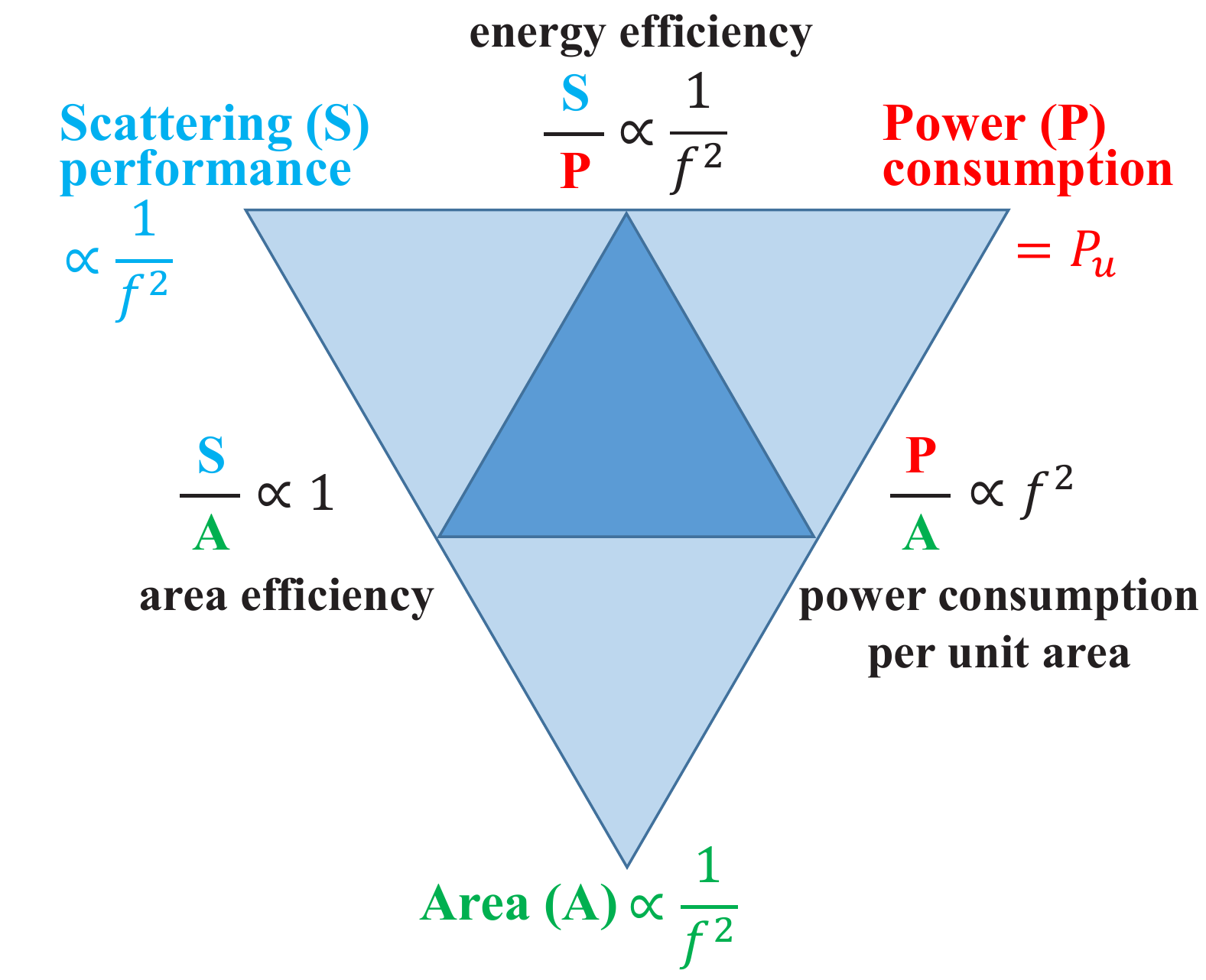}
\caption{SPA triangle of the unit cell of an RIS.}
\label{PPA}
\end{figure}

Specifically, (\ref{s33}) indicates that the intensity of $s_{n,m}^r$ is proportional to the area of the unit cell. In other words, the scattering performance of a unit cell is proportional to ${{d_x}{d_y}}$. Since $d_x$ and $d_y$ are usually a fraction of the wavelength scale, the scattering performance of a unit cell is thus inversely proportional to $f^2$, where $f$ denotes the operating frequency of the RIS. As far as the average power consumption ($P_u$) of a unit cell is concerned, it is independent of the operating frequency and is often of the order of the mW. For example, $P_u$ is equal to 2.8 mW, 6.2 mW, and 3.75 mW for the RISs reported in \cite{pathloss} and \cite{mmWaveRIS6W} that operate in different frequency bands. Specifically, $P_u$ accounts for the internal tunable components and the external control circuits. On the other hand, the area of a unit cell is simply ${{d_x}{d_y}}$ and it is hence inversely proportional to $f^2$. The relation between the scattering performance, the power consumption, and the area of a unit cell, along with the corresponding efficiencies, is illustrated in Fig. \ref{PPA}.

More precisely, the energy efficiency is obtained by computing the ratio between the scattering performance and the power consumption. From Fig. \ref{PPA}, we see that the energy efficiency of a unit cell is inversely proportional to $f^2$. The area efficiency is defined as the ratio between the scattering performance and the area of the unit cell. Since both metrics are inversely proportional to $f^2$, the area efficiency is independent of the operating frequency. Finally, the power consumption per unit area is proportional to $f^2$.

The metrics and efficiencies reported in Fig. \ref{PPA} allow us to better understand and quantify the performance of RISs with different operating frequencies. As an example, let us consider two RISs that operate at 3 GHz and 30 GHz. Based on Fig. \ref{PPA}, the energy efficiency of a single unit cell that operates at 3 GHz is 100 times higher than that of a unit cell that operates at 30 GHz. This implies that, under the same conditions (e.g., the same antenna gains $G_t$ and $G_t$, etc.), an RIS that operates at 30 GHz consumes 100 times more power than an RIS that operates at 3 GHz to achieve the same performance. The basic reason is that the RIS at 30 GHz needs 100 times more unit cells than the RIS at 3 GHz to obtain the same performance, since the scattering performance of a single unit cell of the former is 100 times lower than that of the latter. This result can also be directly validated by (\ref{s28}) even though it is a special case.


It can be concluded from the above example that the deployment of RISs in high frequency bands needs a careful system design that accounts for the tradeoff between performance and power consumption. According to (\ref{s33}), one potential solution is to increase the transmit/receive antenna gains $G_t$ and $G_r$, which can be realized by employing high gain antennas or large antenna arrays at the transceivers. In addition, another potential solution is to use fixed-coded RIS for specific applications in high frequency bands like coverage expansion. This type of RISs (i.e., the fixed-coded metasurfaces introduced in \cite{MetaCoding} and employed in \cite{docomo}) is completely passive, and its power consumption is zero and hardware cost is extremely low even if its area is large.

Finally, we discuss the impact of the angle-dependent factor ${F_{n,m}^{combine}}$ on the path loss models. ${F_{n,m}^{combine}}$ can be maximized to 1 only when the angles $\theta_{n,m}^{tx}$, $\theta_{n,m}^t$, $\theta_{n,m}^r$, and $\theta_{n,m}^{rx}$ in (\ref{s23}) are all equal to 0 degrees. As a result, the factor ${F_{n,m}^{combine}}$ ensures that, as the size of the RIS increases to infinity, the received power from the entire RIS is still finite \cite{vector}. Last but not least, ${F_{n,m}^{combine}}$ is transmitter-receiver reciprocal, which ensures that the proposed path loss models are reciprocal, as long as the reflection coefficient ${\varGamma _{n,m}}$ is reciprocal for angles of incidence and reception.

\section{Measurements in the mmWave Band to Validate the Path Loss Models}\label{MeasurementandDiscussion}
In this section, we report the experimental measurements in the mmWave frequency band to validate the free-space path loss models for RIS-assisted wireless communications.
\subsection{Experimental Setup and Fabricated RISs}\label{SystemSetup}
In the following, we introduce and describe the fabricated RISs that are employed in the measurements, as well as the experimental setup and calibration method.
\subsubsection{Two Fabricated mmWave RISs}\label{RISSamples}
\begin{figure}
\centering
\includegraphics[scale = 0.32]{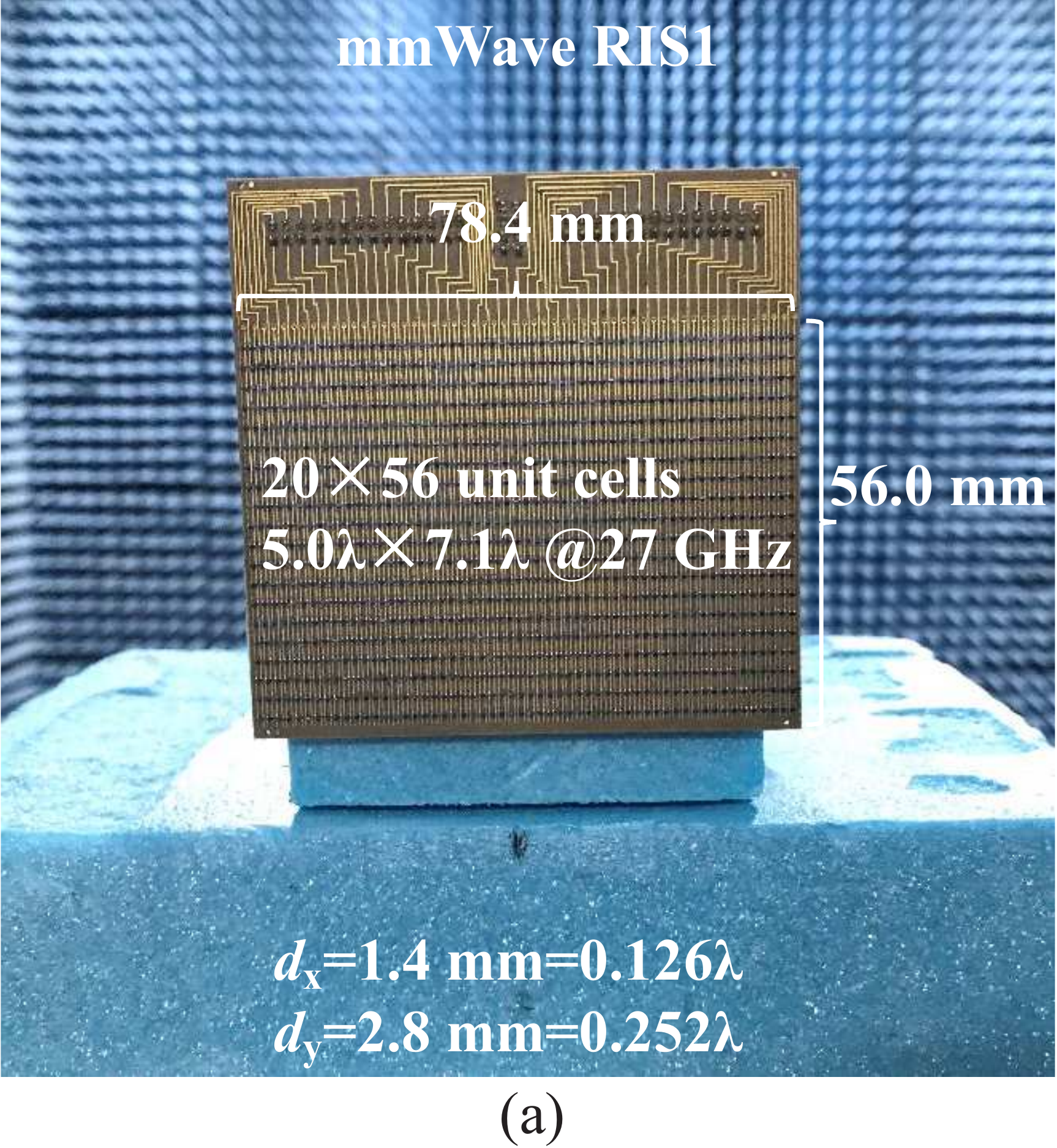}
\hspace{0.6cm}
\includegraphics[scale = 0.32]{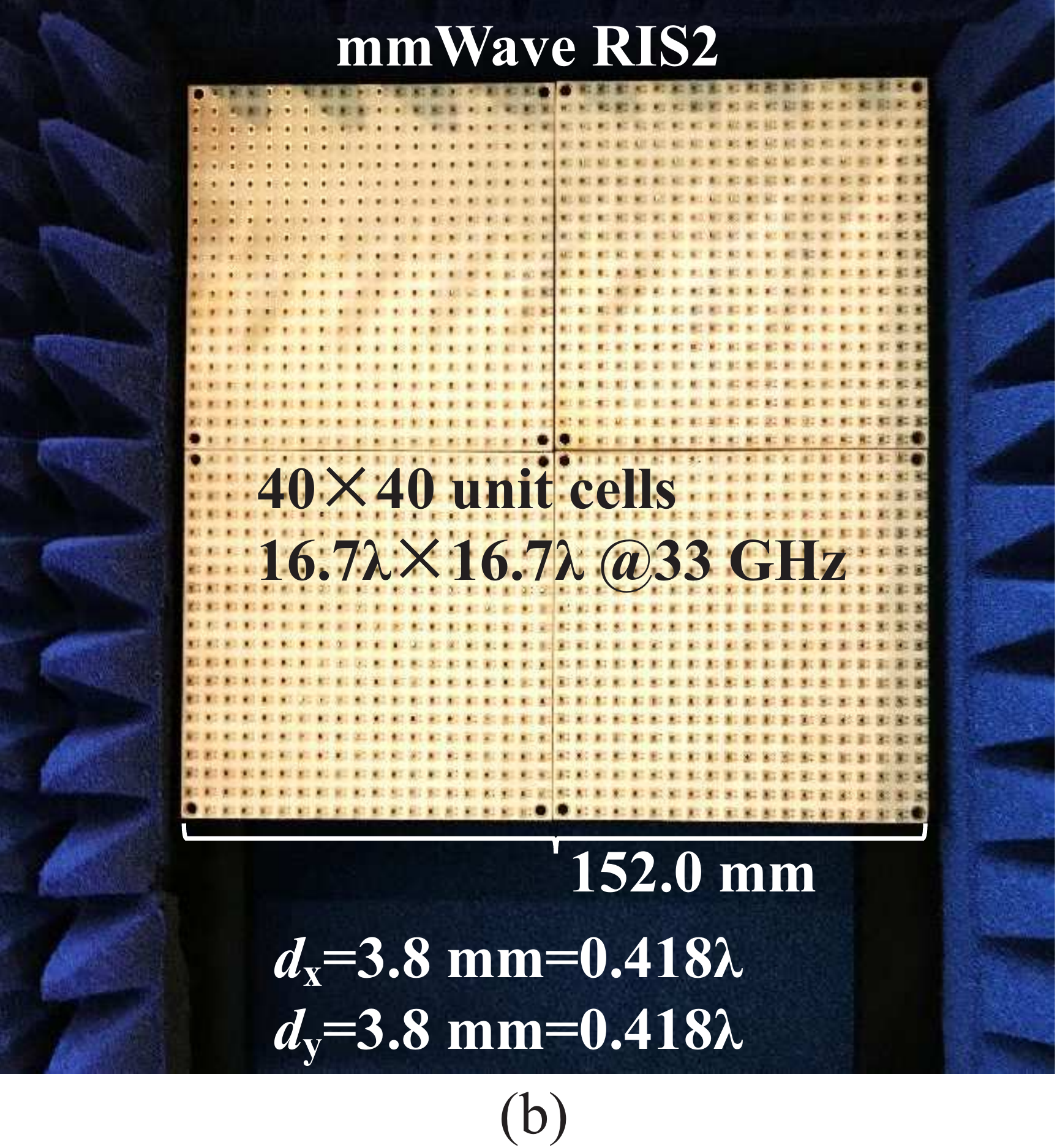}
\caption{Photographs of the two mmWave metasurfaces utilized in the measurements. (a) mmWave RIS1. (b) mmWave RIS2.}
\label{RISphotos}
\end{figure}

Fig. \ref{RISphotos} shows the two fabricated metasurfaces that are utilized to conduct the path loss measurements in the mmWave frequency band. For ease of exposition, the metasurfaces in Fig. \ref{RISphotos}(a) and Fig. \ref{RISphotos}(b) are referred to as mmWave RIS1 and mmWave RIS2, respectively. Both metasurfaces are phase-programmable with 1-bit coding. The detailed parameters of the RISs are reported in Table \ref{parasummary}.
\subsubsection{Measurement Systems and Setup}\label{MeasurementSystem}
\begin{figure}
\centering
\includegraphics[scale = 0.36]{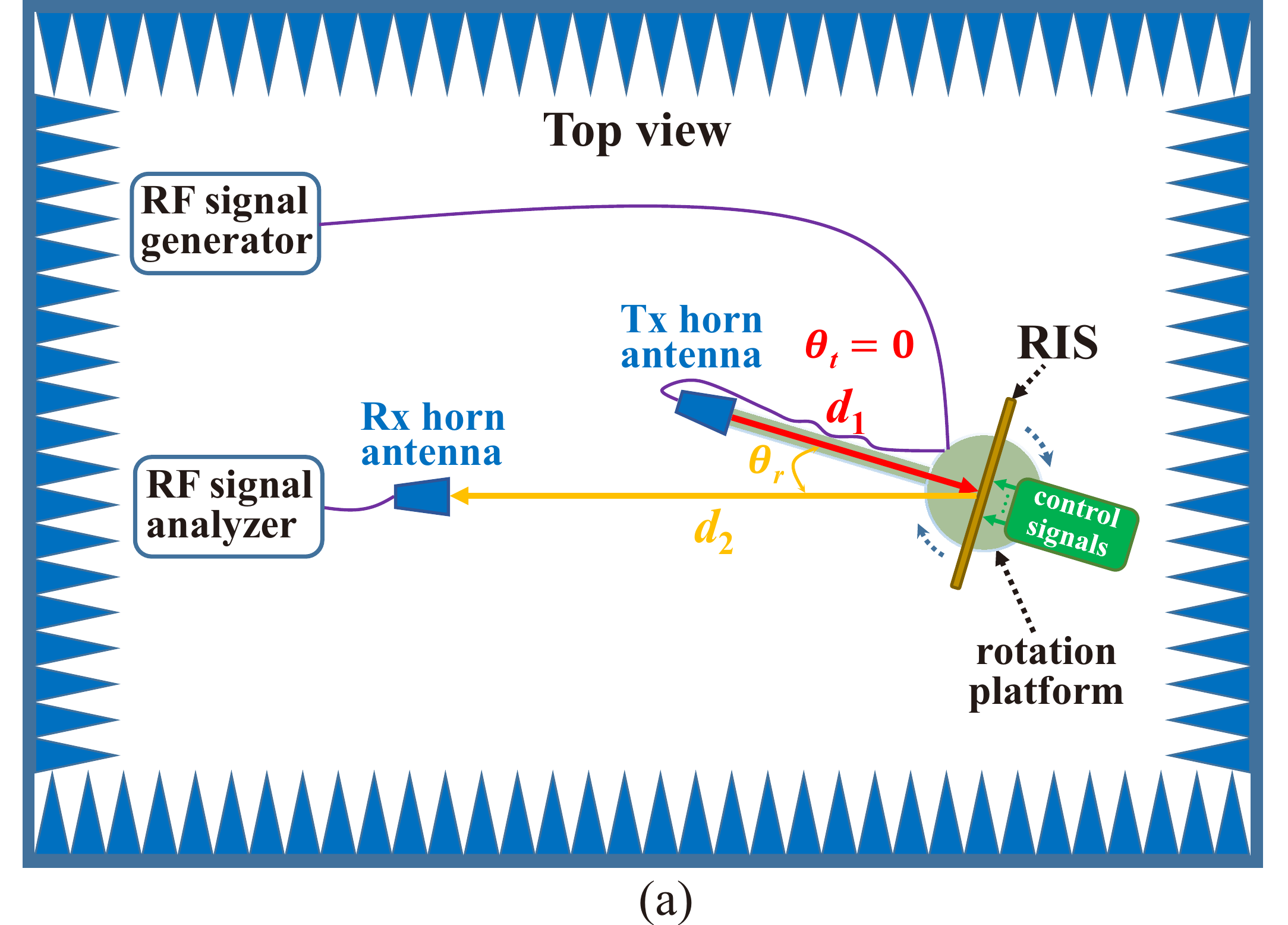}
\hspace{-0.1cm}
\includegraphics[scale = 0.36]{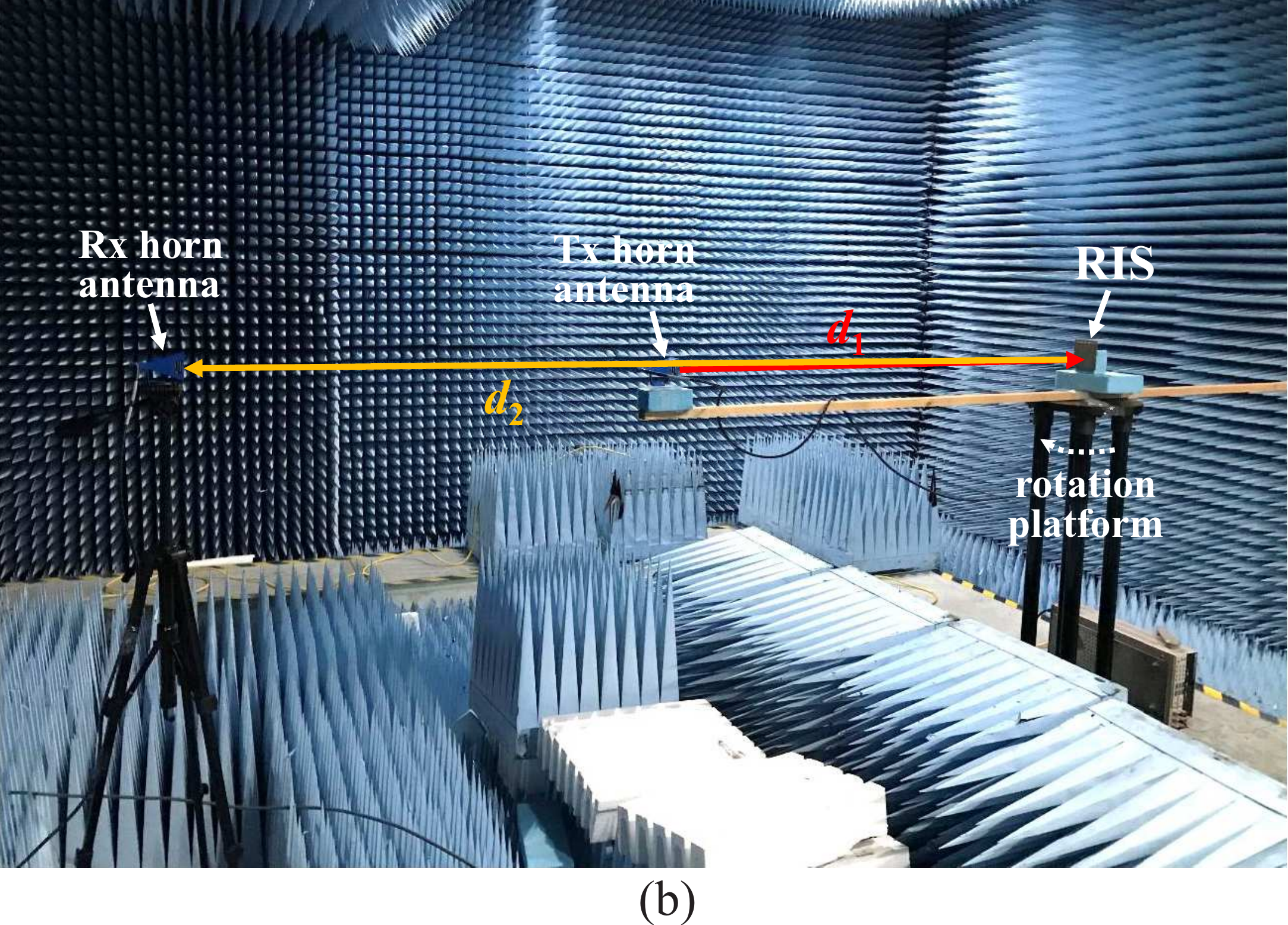}
\caption{Measurement system A: free-space path loss measurement system for measuring the amount of power reflected from the RIS as a function of the angle of reception $\theta_{r}$. (a) Diagram. (b) Photograph.}
\label{MeasureSystems}
\end{figure}

\begin{table}
\centering
\footnotesize
\caption{Specifications of the mmWave RISs and Antennas Utilized in the Measurements.}\label{parasummary}
\begin{tabular}{|c|c|}
\hline
\textbf{Name} & \textbf{Parameters}  \\
\hline
mmWave RIS1 & \tabincell{l}{$N = 20$, $M = 56$, $d_{x}=1.4\ $mm, $d_{y}=2.8\ $mm, operating frequency $f=27\ $GHz, 1-bit phase control.\\ $\angle {\varGamma _{n,m}} = 165^\circ$ and $\left| {\varGamma _{n,m}} \right|=0.9$ when coding ``0'' is applied to $\rm{U}_{n,m}$, $\angle {\varGamma _{n,m}} = 0^\circ$ and $\left| {\varGamma _{n,m}} \right|=0.7$\\ when coding ``1'' is applied to $\rm{U}_{n,m}$. The Fraunhofer distance boundary between the far field and the\\ near field of the mmWave RIS1 is $\frac{2{D^2}}{\lambda}=1\ $m.}\\
\hline
mmWave RIS2& \tabincell{l}{$N = 40$, $M = 40$, $d_{x}=3.8\ $mm, $d_{y}=3.8\ $mm, operating frequency $f=33\ $GHz, 1-bit phase control.\\ $\angle {\varGamma _{n,m}} = 150^\circ$ and $\left| {\varGamma _{n,m}} \right|=0.8$ when coding ``0'' is applied to $\rm{U}_{n,m}$, $\angle {\varGamma _{n,m}} = 0^\circ$ and $\left| {\varGamma _{n,m}} \right|=0.8$\\ when coding ``1'' is applied to $\rm{U}_{n,m}$. The Fraunhofer distance boundary between the far field and the\\ near field of the mmWave RIS2 is $\frac{2{D^2}}{\lambda}=5\ $m.}\\
\hline
\tabincell{c}{mmWave\\ horn antenna}& \tabincell{l}{When the operating frequency is $f{=}27\ $GHz, $G_t=G_r=109.6=20.4\ $dB. According to (\ref{s11}) and (\ref{s12}),\\ $F^{tx}(\theta ,\varphi ) = F^{rx}(\theta ,\varphi ) = \left({\cos \theta }\right)^{53.8} $ when $\theta  \in \left[ {0,\frac{\pi }{2}} \right]$, $F^{tx}(\theta ,\varphi ) = F^{rx}(\theta ,\varphi ) = 0$ when $\theta  \in \left( {\frac{\pi }{2},\pi } \right]$.\\
When the operating frequency is $f{=}33\ $GHz, $G_t=G_r=128.8=21.1\ $dB. According to (\ref{s11}) and (\ref{s12}),\\ $F^{tx}(\theta ,\varphi ) = F^{rx}(\theta ,\varphi ) = \left({\cos \theta }\right)^{63.4} $ when $\theta  \in \left[ {0,\frac{\pi }{2}} \right]$, $F^{tx}(\theta ,\varphi ) = F^{rx}(\theta ,\varphi ) = 0$ when $\theta  \in \left( {\frac{\pi }{2},\pi } \right]$.}\\
\hline
\end{tabular}
\end{table}
To validate the proposed free-space path loss model for RIS-assisted communications, we have built two different measurement systems, which are referred to as measurement system A and B in further text. The measurement system A, which is shown in Fig. \ref{MeasureSystems}, is widely used for measuring the radiation patterns of antennas in microwave engineering. We use this system to measure the power distribution of the signal reflected from the RIS as a function of the angle of reception. As shown in Fig. \ref{MeasureSystems}, the Tx horn antenna and the RIS are fixed on the rotation platform, and the incident signal is perpendicular to the RIS, i.e., $\theta_{t}=0$ (normal incidence). When the platform is rotated, the receiver, which is constituted by an Rx horn antenna and an RF signal analyzer, can measure the received power at different angles of reception $\theta_{r}$.
The Tx and Rx horn antennas are identical and the corresponding specifications are reported in Table \ref{parasummary}. The main advantage of the measurement system A in Fig. \ref{MeasureSystems} lies in the possibility of easily measuring the distribution of the power reflected from the RIS as a function of the observation (reception) angle and in comparing the resulting measurements with (\ref{s29}).

\begin{figure}
\centering
\includegraphics[scale = 0.36]{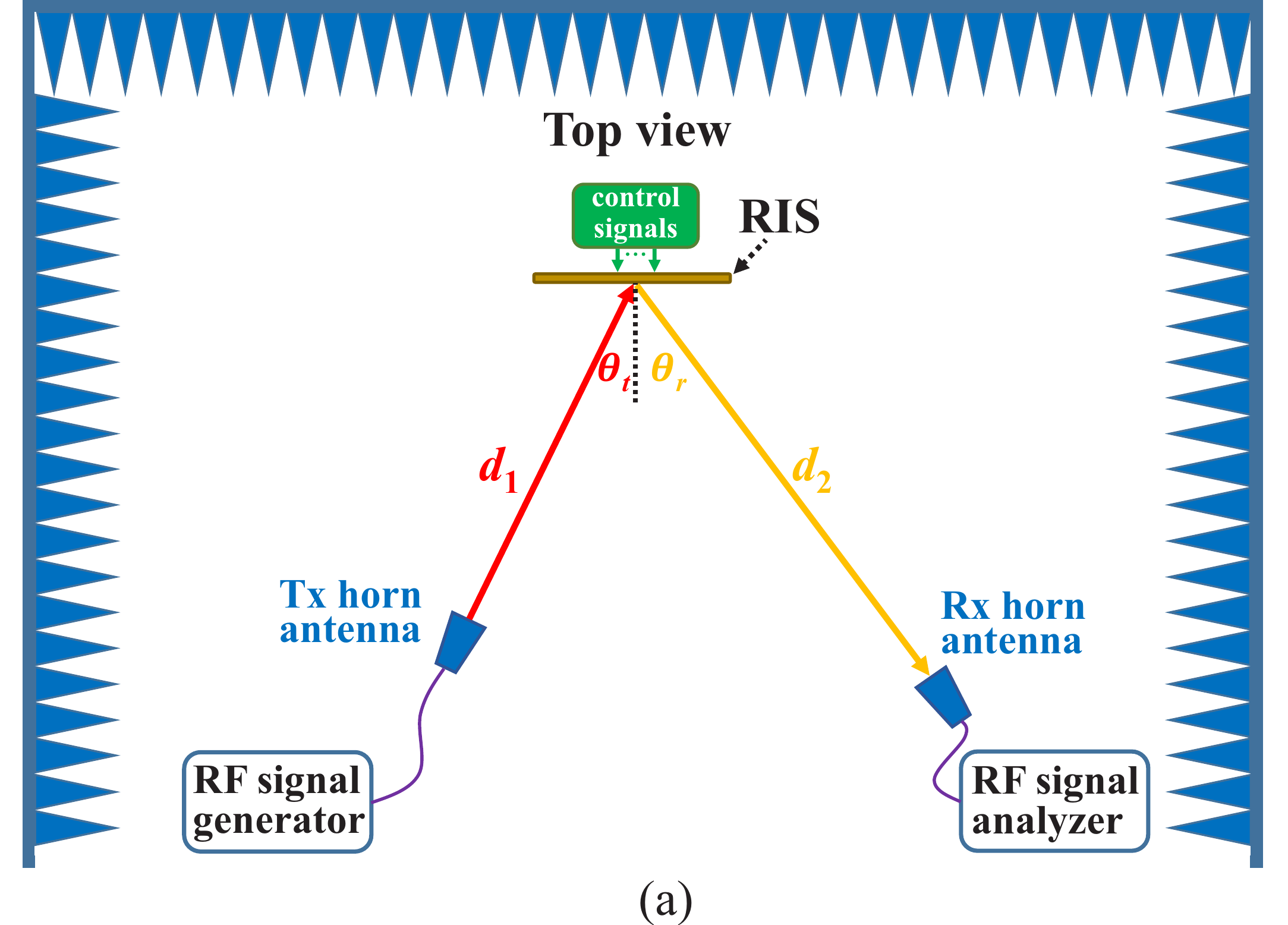}
\hspace{-0.1cm}
\includegraphics[scale = 0.36]{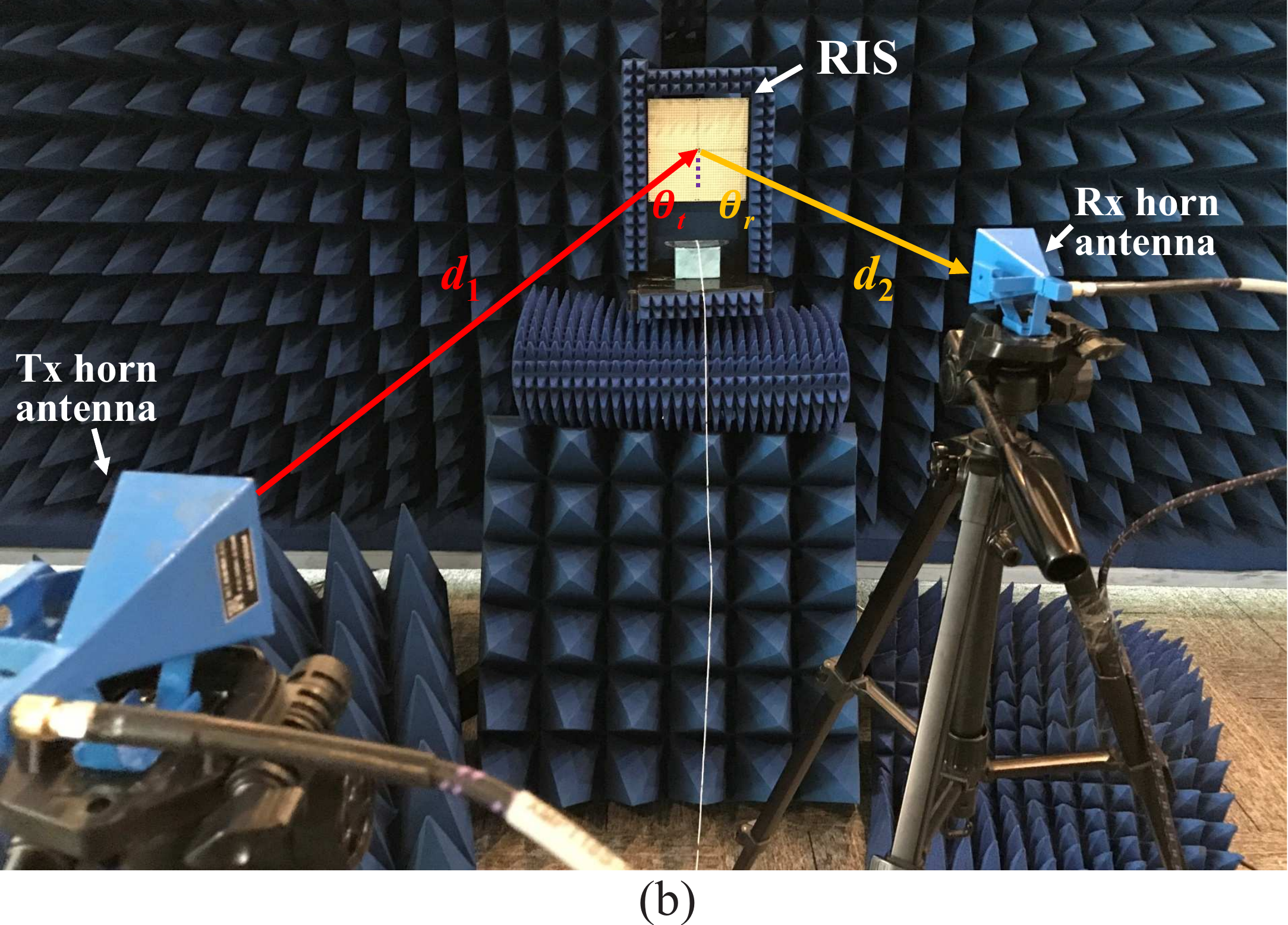}
\caption{Measurement system B: free-space path loss measurement system for measuring the amount of power reflected from the RIS for different configurations of $d_{1}$, $d_{2}$, $\theta_{t}$, and $\theta_{r}$. (a) Diagram. (b) Photograph.}
\label{MeasureSystems2}
\end{figure}

The main limitation of the measurement system A lies, on the other hand, in the fact that the Tx horn antenna is kept fixed on the rotation platform. As a result, the distance $d_1$ cannot be flexibly changed during the measurements. In order to study the validity of the proposed path loss model as a function of the transmission distances $d_1$ and $d_2$, therefore, we have built the measurement system B that has been deployed in a large empty room by using movable walls coated with absorbing materials, as shown in Fig. \ref{MeasureSystems2}. The components of the measurement systems A and B are the same. The only difference is that no rotation platform is utilized in the measurement system B. The RIS is, on the other hand, placed on a stable tripod, so that the transmit and receive antennas can be flexibly moved in order to facilitate the measurements for different configurations of $d_{1}$, $d_{2}$, $\theta_{t}$, and $\theta_{r}$.

\subsubsection{Calibration Method}\label{Calibration}
\begin{figure}
\centering
\includegraphics[scale = 0.4]{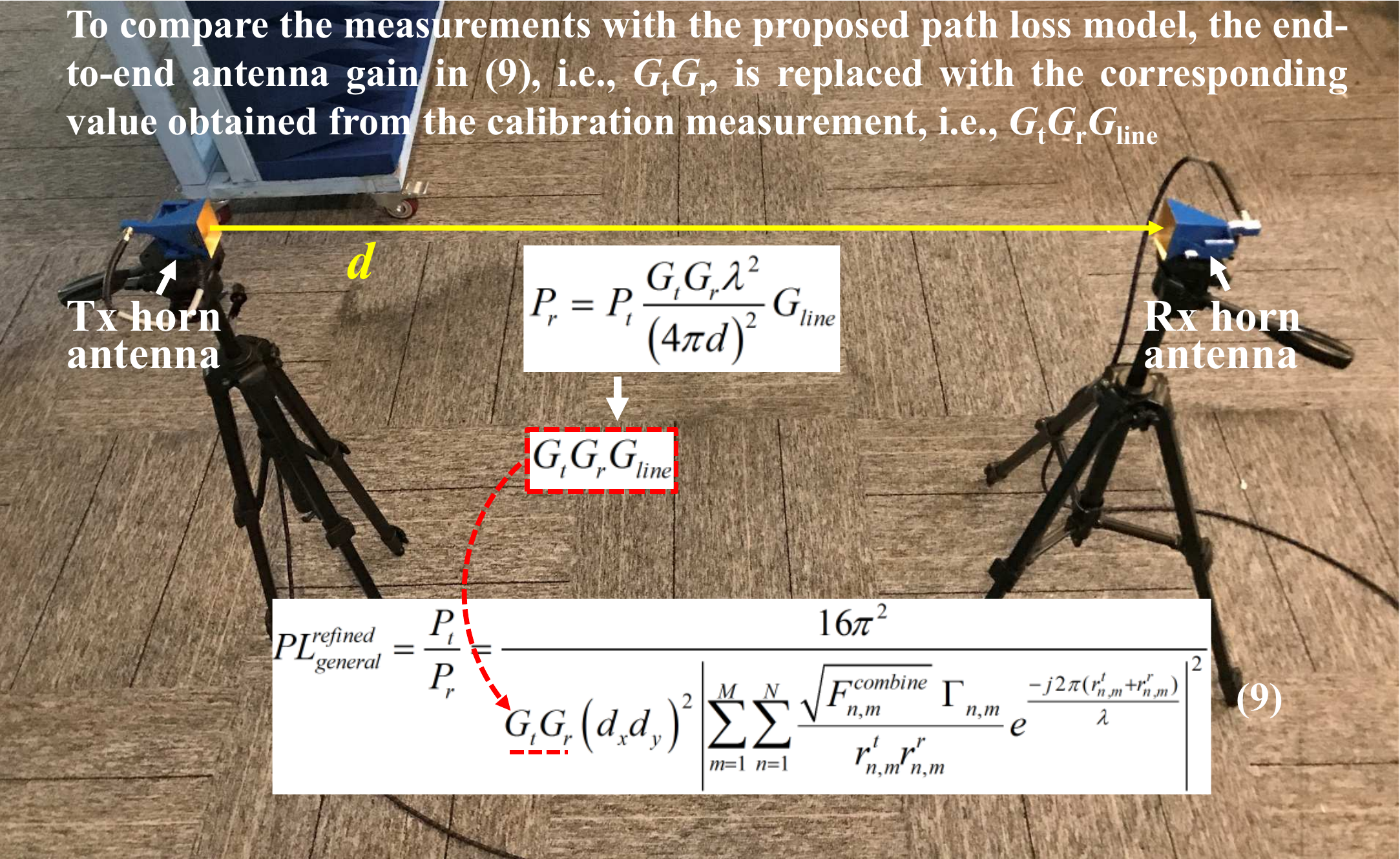}
\caption{Illustration of the calibration method.}
\label{calibration}
\end{figure}
As shown in Fig. \ref{MeasureSystems} and Fig. \ref{MeasureSystems2}, several RF cables are necessary for connecting the horn antennas and measurement instruments, which introduce extra power loss to the transmitted and received signals. In addition, the gain of the horn antennas slightly varies at different operating frequencies. In order to appropriately compare the measurement results with the proposed path loss model, we employ the calibration method illustrated in Fig. \ref{calibration}. As shown in Fig. \ref{calibration}, by aligning the transmit antenna with the receive antenna, and letting the transmitted signal propagate in free space, the received power can be written as\cite{Book}
\begin{equation}\label{s32}
{P_r} = {P_t}\frac{{G_t}{G_r}{\lambda ^2}}{{\left({4\pi d}\right)}^2}{G_{line}},
\end{equation}
where $G_{line}$ denotes the gain (i.e., the reciprocal of the loss) introduced by the RF cables, which is a negative value in dB. Based on (\ref{s32}), we have measured the actual value of $G_tG_rG_{line}$ through the calibration setup illustrated in Fig. \ref{calibration} for each considered measurement system and operating frequency. The resulting calibration parameters are listed in Table \ref{CalibrationPara}. As described in Fig. \ref{calibration}, the measured values of $G_t G_r G_{line}$ are used in lieu of $G_t G_r$ in (\ref{s29}). In further text, all theoretical curves obtained with (\ref{s29}) are computed by using the calibration parameters in Table \ref{CalibrationPara}. This applies also to the curves that correspond to the measurements reported in \cite{pathloss}.
\begin{table}
\centering
\footnotesize
\caption{Measured Calibration Parameters.}\label{CalibrationPara}
\begin{tabular}{|c|c|c|}
\hline
\textbf{System Name} & \textbf{Operating Frequency}  & \textbf{Measured Calibration Parameter}\\
\hline
Measurement system A & $f=27\ $GHz & $G_tG_rG_{line}=2.9\ $dB\\
\hline
Measurement system B & $f=27\ $GHz & $G_tG_rG_{line}=24.0\ $dB\\
\hline
Measurement system B & $f=33\ $GHz & $G_tG_rG_{line}=22.0\ $dB\\
\hline
\end{tabular}
\end{table}
\subsection{Comparison with a Rectangular Metal Plate}\label{ComparisonwithMetal}
\begin{figure}
\centering
\includegraphics[scale = 0.4]{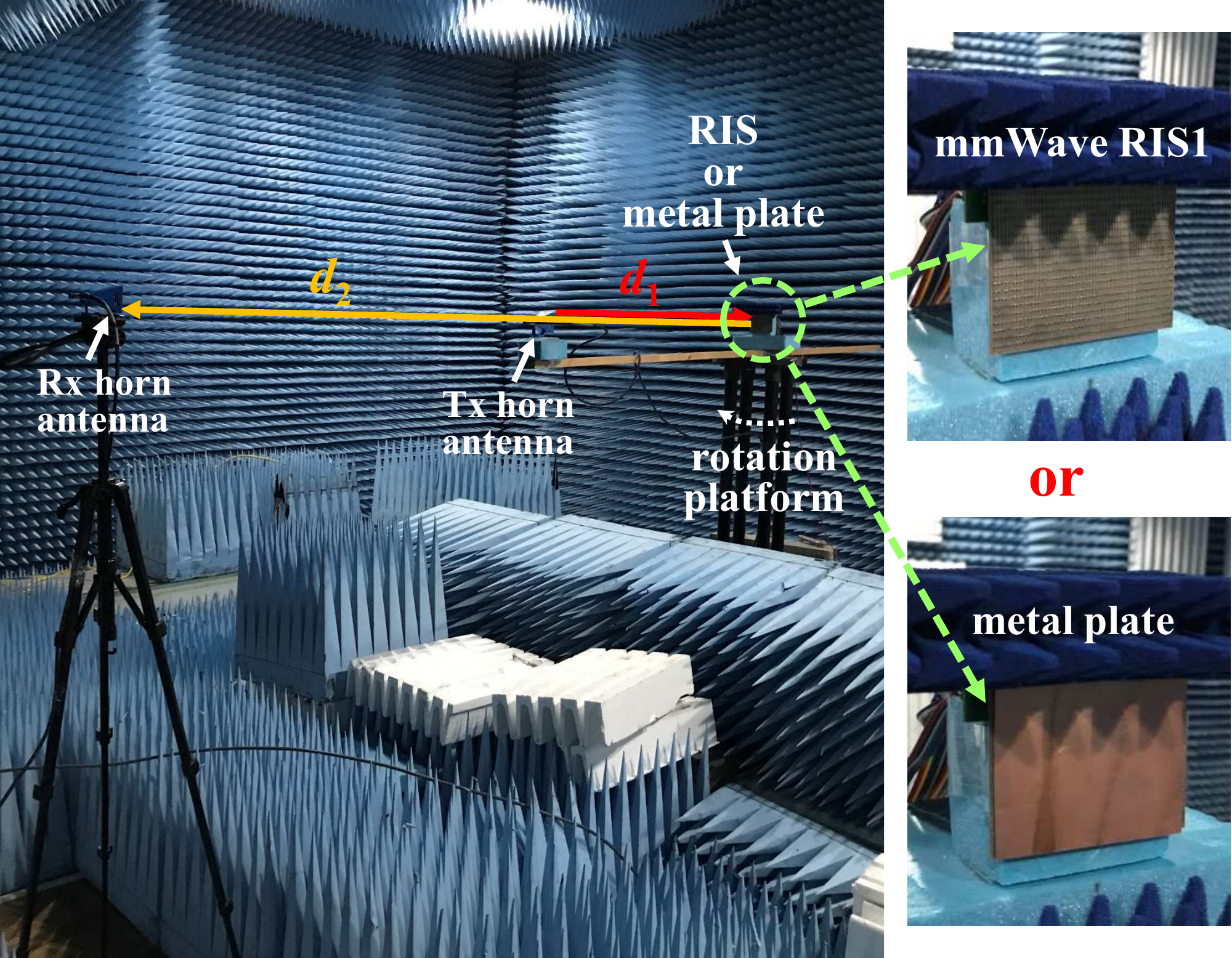}
\caption{Comparison of the reflection properties between the mmWave RIS1 and a metal plate with the same shape and size. The measurement system A is used.}
\label{comparisonmetal}
\end{figure}

In Section \ref{SectionIII}, we have benchmarked the power scattered by an RIS with that of a rectangular metal plate of the same size and shape as the RIS.
In this subsection, we describe the measurement setup in order to experimentally validate whether a uniformly configured RIS can mimic the scattering from an ideal metal plate, e.g., if the scattered power is the same. The considered setup is illustrated in Fig. \ref{comparisonmetal}, in which a metal plate of the same size and shape as the mmWave RIS1 is considered. The experiments are conducted by using the measurement system A. More precisely, the measurement setup is $P_t=20\ $dBm, $f=27\ $GHz, $\theta_{t}=0$, $d_{1}=1.3\ $m, and $d_{2}=2.6\ $m. As far as the mmWave RIS1 is concerned, all of its unit cells are configured to the coding state ``0'', which implies $\left| {\varGamma _{n,m}} \right|=0.9$ and $\angle {\Gamma _{n,m}} = {165^\circ}$ according to Table \ref{parasummary}. As far as the metal plate is concerned, we cover the mmWave RIS1 with the metal plate in order to ensure that all measurement settings remain unchanged, as shown in Fig. \ref{comparisonmetal}.
\begin{figure}
\centering
\includegraphics[scale = 0.46]{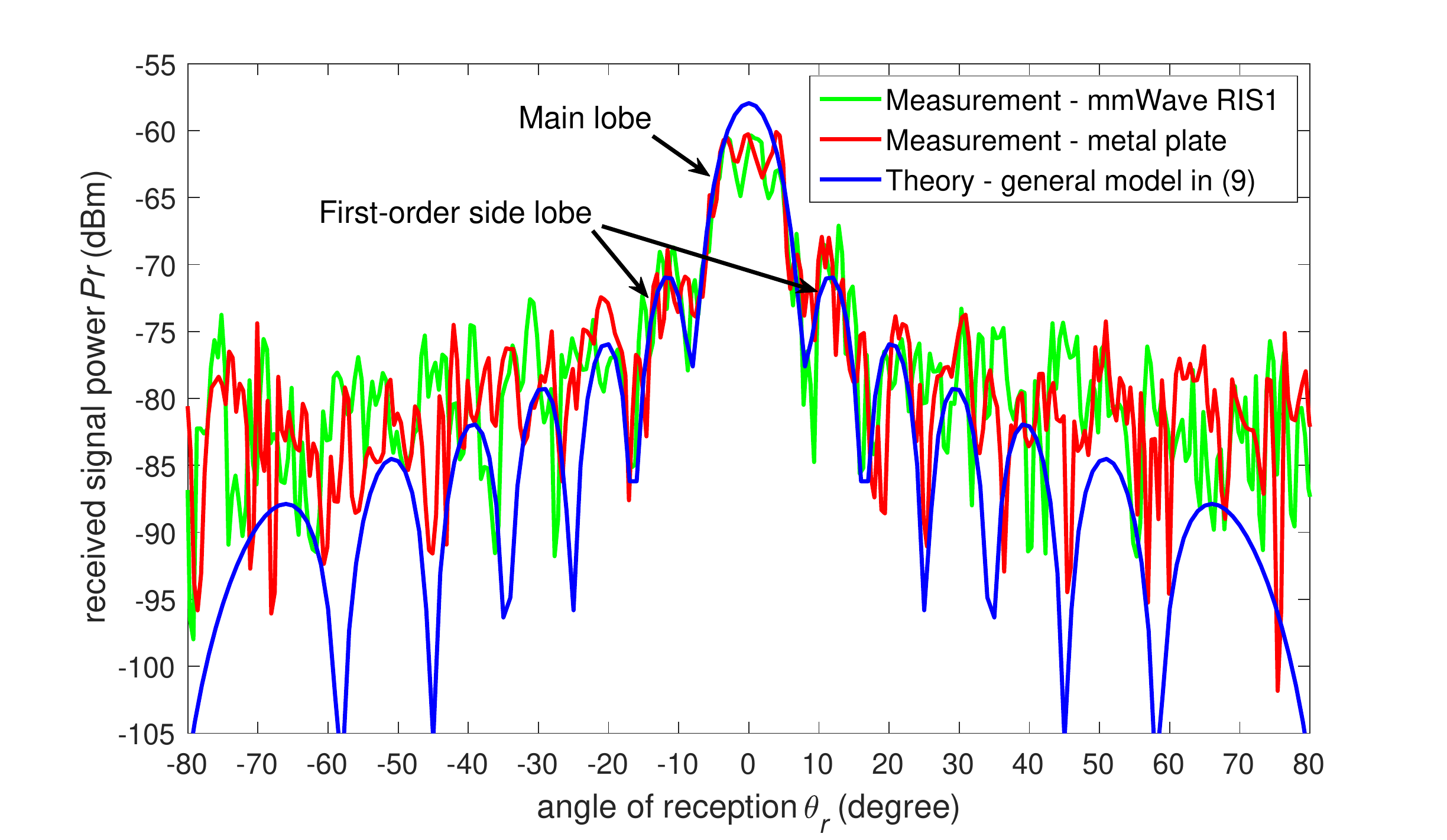}
\caption{Measurement results of the reflected signal power for the mmWave RIS1 and a metal plate of the same shape and size when $P_t=20\ $dBm, $\theta_{t}=0$, $d_{1}=1.3\ $m, and $d_{2}=2.6\ $m.}
\label{comparisonmetalresult}
\end{figure}

The measurement results illustrated in Fig. \ref{comparisonmetalresult} prove that an RIS that is configured with a uniform coding state and with a high reflection efficiency has the same reflection characteristics as a metal plate of the same shape and size. In particular, we observe that the measured main lobe and the first-order side lobe of the reflected signal power from the RIS and the metal plate are in good agreement\footnote{For ease of illustration, the angle of reception $\theta_r$ is negative when the receive antenna is located in the left-hand side of the RIS, i.e., $\varphi_r=\pi$. When the receive antenna is located in the right-hand side of the RIS, i.e., $\varphi_r=0$, $\theta_r$ is greater than zero.}.
When the angle of reception $\theta_r$ is around 0 degree, the received link is blocked by the transmit antenna, which results in an attenuation and fluctuation of the received power in the main lobe. Furthermore, the power scattered by the mmWave RIS1 is slightly lower than that of the metal plate in the main lobe, because the amplitude of the reflection coefficient of its unit cells is 0.9, which is less than that of a metal plate. By taking into account these practical considerations, the experimental results reported in Fig. \ref{comparisonmetalresult} are in good agreement with the proposed path loss model in (\ref{s29}). This also confirms that it is meaningful to validate the relation between the scattering gain $G$ of the unit cell and its size $d_xd_y$ by using a metal plate as benchmark.
\subsection{Normalized Power Radiation Pattern of the Unit Cell}\label{PatternMeasurement}
\begin{figure}
\centering
\includegraphics[scale = 0.6]{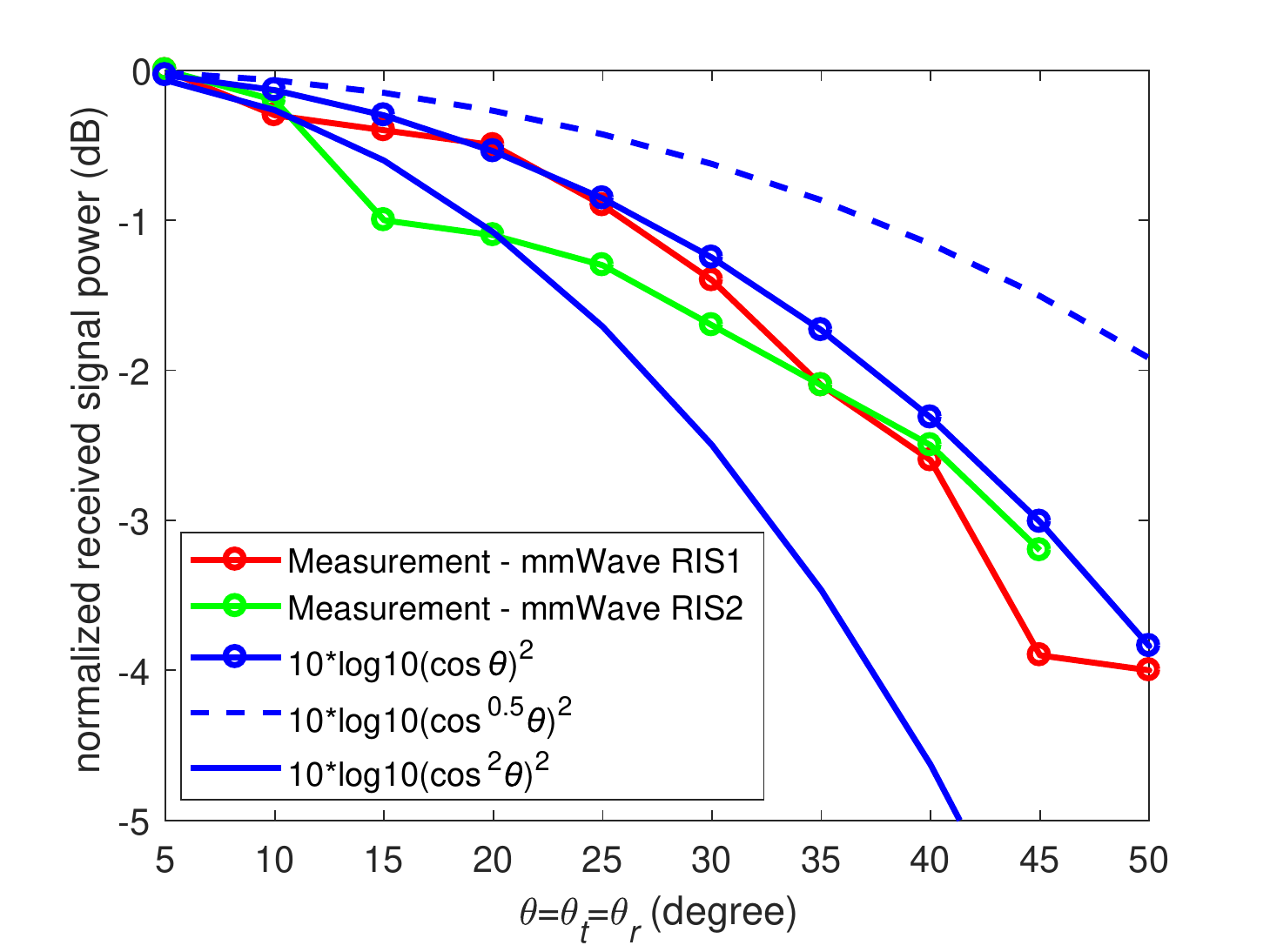}
\caption{Experimental validation of the normalized power radiation pattern of a single unit cell for $\theta_{t}=\theta_{r}=\theta$ in the far-field beamforming setup.}
\label{patternresult}
\end{figure}
The proposed path loss model for RIS-assisted communications is derived by assuming that the normalized power radiation pattern of each unit cell is modeled as given in (\ref{s20}), i.e., the shape of the normalized power radiation pattern is $\cos\theta$.
We measure the received power of the RIS as a whole and test the predicted path loss model in (\ref{s28}) for different functions for the power radiation pattern of a unit cell. We utilize the measurement system B in Fig. \ref{MeasureSystems2}, and have conducted experimental measurements by using the mmWave RIS1 and mmWave RIS2. The results are illustrated in Fig. \ref{patternresult}.

As far as the mmWave RIS1 is concerned, the measurement setup is $P_t=20\ $dBm, $f=27\ $GHz, $\theta_{t}=\theta_{r}=\theta$, $d_{1}=d_{2}=2\ $m, and all the unit cells are configured to operate in the coding state ``0''. As far as the mmWave RIS2 is concerned, the only differences in the measurement setup are $f=33\ $GHz and $d_{1}=d_{2}=5\ $m. Since the Fraunhofer distance $\frac{2{D^2}}{\lambda}$ is equal to 1 m and 5 m for the mmWave RIS1 and mmWave RIS2, respectively, the measurements are conducted in the far-field region of both RISs. In this setup, both RISs are configured to realize far-field beamforming. According to (\ref{s28}), the received signal power is proportional to the square of the normalized power radiation pattern of each unit cell, i.e., $\cos {\theta _t}\cos {\theta _r} = {\left( {\cos {\theta}} \right)^2}$. The experimental measurements illustrated in Fig. \ref{patternresult} confirm this modeling assumption. 
\subsection{Validation via Measurement Results by Using the mmWave RIS1}\label{MeasurementmmWaveRIS1}
In this subsection, we validate the proposed path loss model by using the mmWave RIS1, when it is configured for specular reflection and for intelligent (anomalous) reflection. In specular reflection, all the unit cells of the RIS are configured to be in the same coding state, i.e., the RIS is an equiphase surface. In this setup, the angle of reflection is the same as the angle of incidence\cite{pathloss}.
In contrast, the coding states of the unit cells of the RIS need to be carefully designed to realize intelligent reflection, e.g., in order to obtain an angle of reflection that is different from the angle of incidence. An example is constituted by an RIS that realizes dual-beam intelligent reflection, which needs a stripe coding pattern for the unit cells\cite{MetaCoding,MetaInfo}.
\subsubsection{Specular Reflection via the mmWave RIS1}\label{SpecularmmWaveRIS1}
\begin{figure}
\centering
\includegraphics[scale = 0.525]{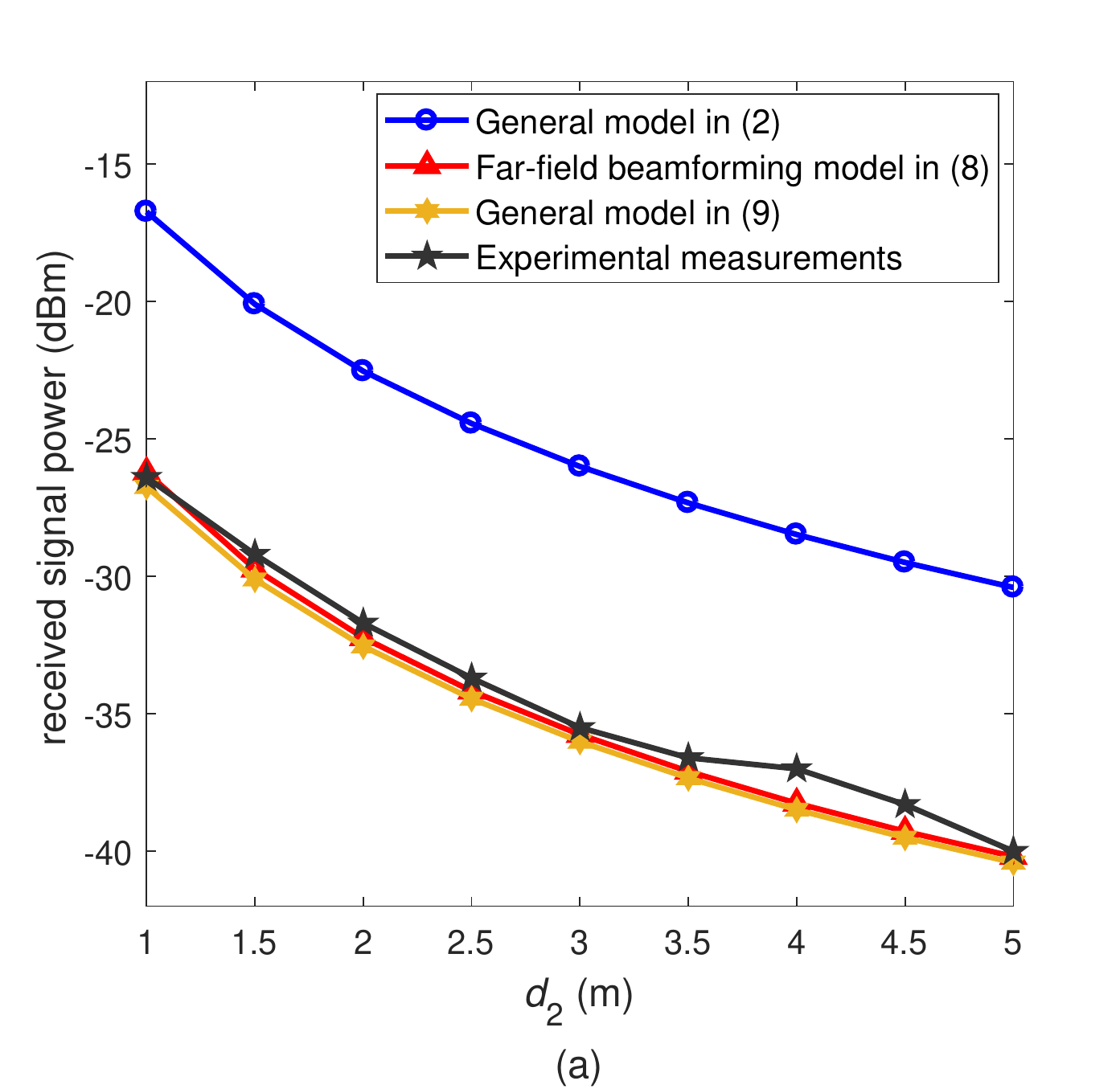}\hspace{-0.5cm}
\includegraphics[scale = 0.525]{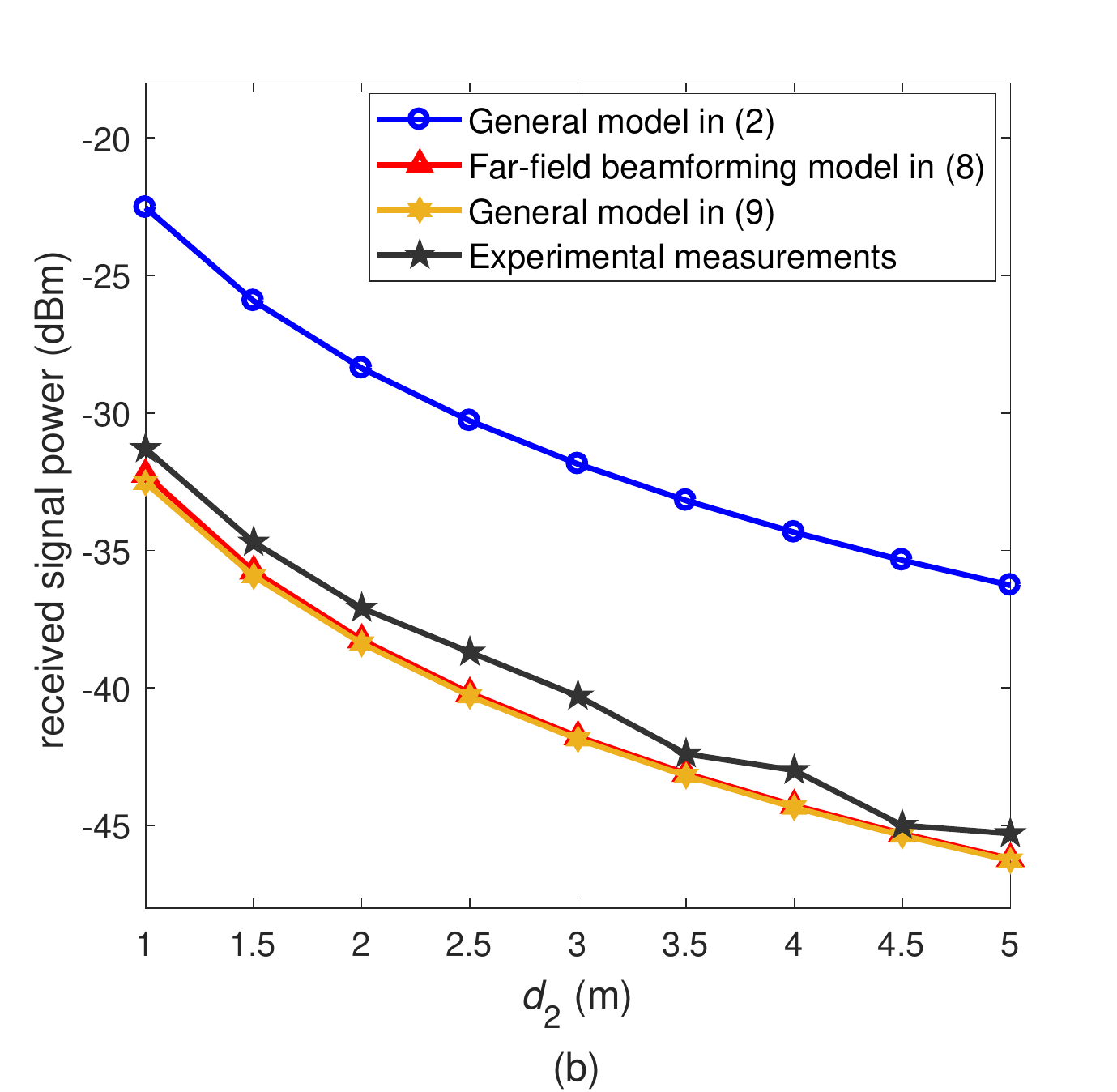}
\includegraphics[scale = 0.525]{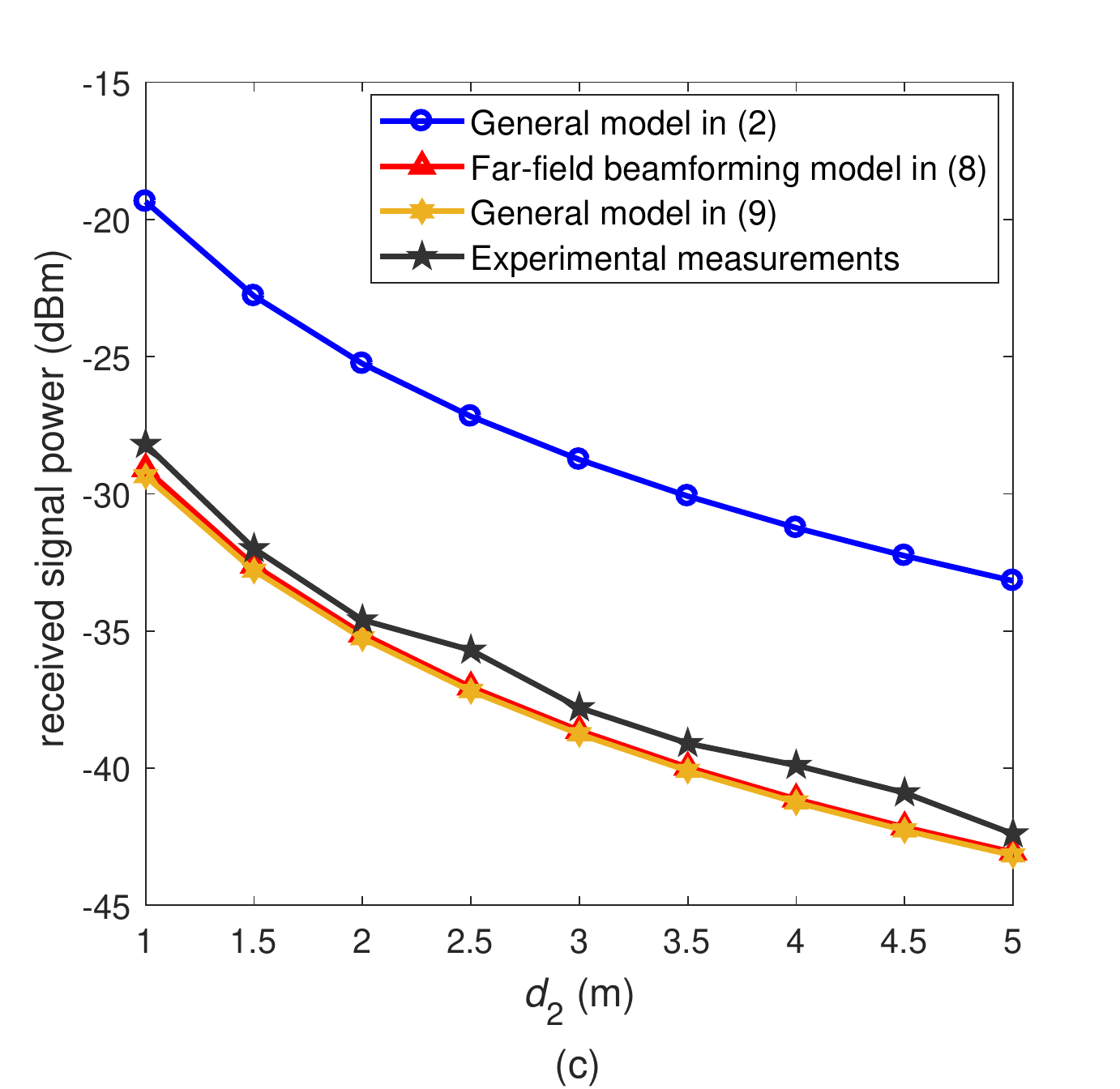}\hspace{-0.5cm}
\includegraphics[scale = 0.525]{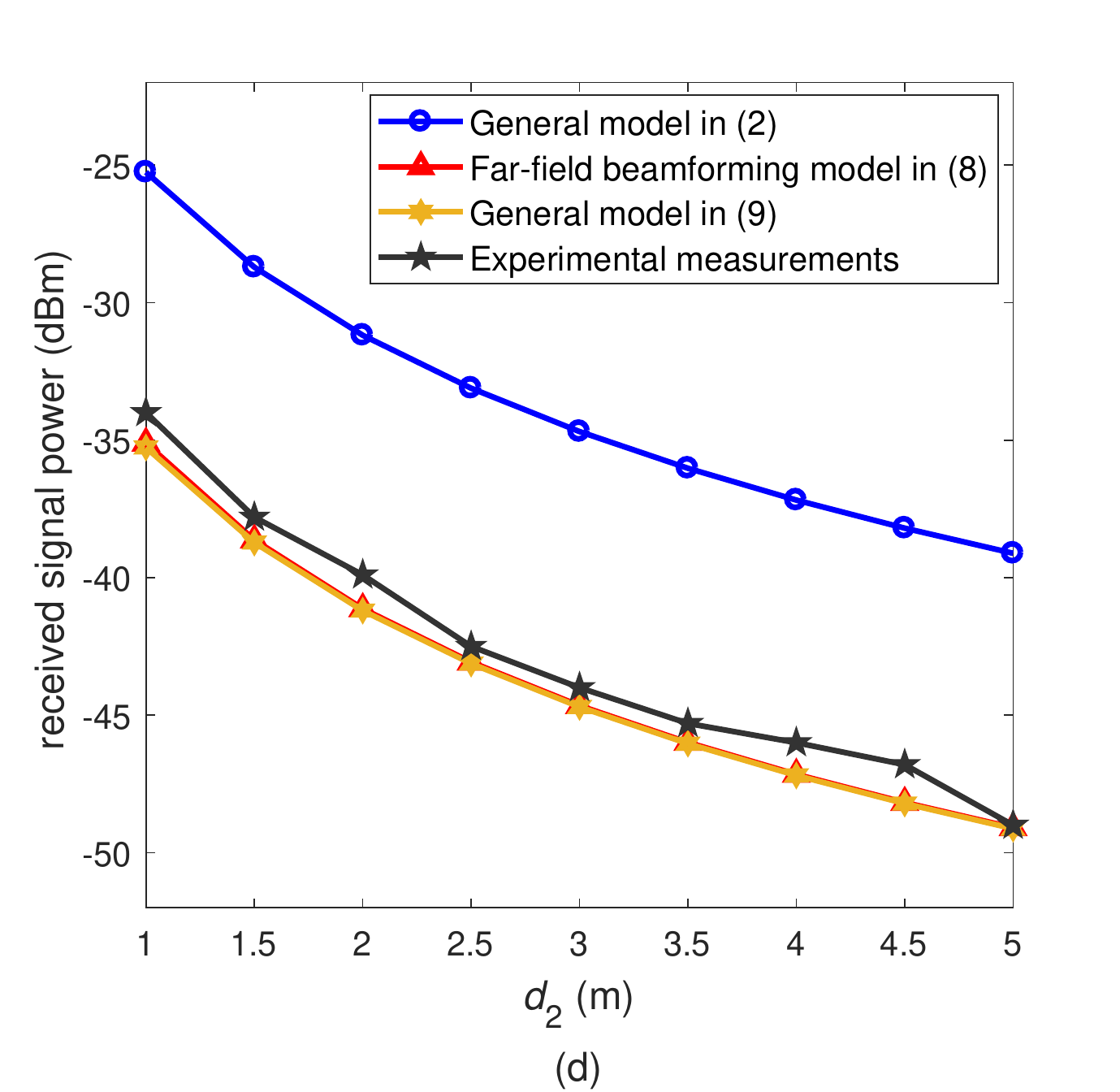}
\caption{Specular reflection via mmWave RIS1: Measurements vs. modeling. (a) Received signal power versus $d_2$ when $d_1=1\ $m and $\theta_{t}=\theta_{r}=10^\circ$. (b) Received signal power versus $d_2$ when $d_1=2\ $m and $\theta_{t}=\theta_{r}=10^\circ$. (c) Received signal power versus $d_2$ when $d_1=1\ $m and $\theta_{t}=\theta_{r}=45^\circ$. (d) Received signal power versus $d_2$ when $d_1=2\ $m and $\theta_{t}=\theta_{r}=45^\circ$.}
\label{mmWaveRIS1mea1}
\end{figure}

The measurement setup is $P_t=20\ $dBm, $f=27\ $GHz, $\theta_{t}=\theta_{r}=10^\circ$, and all the unit cells of the mmWave RIS1 are configured in coding state ``0''. The experiments are conducted by using the measurement system B, so that $d_{1}$ and $d_{2}$ can be flexibly varied during the measurements. By letting $d_{1}=1\ $m or $d_1=2\ $m, and varying $d_{2}$ in the range $[1,5]$ m, we have measured the received power as a function of $d_{1}$ and $d_{2}$. The same experiment is repeated by considering $\theta_{t}=\theta_{r}=45^\circ$. Fig. \ref{mmWaveRIS1mea1} illustrates the measured received signal power as a function of $d_{1}$, $d_{2}$, $\theta_{t}$, and $\theta_{r}$. We observe that the measurement results are in good agreement with the proposed general path loss model given in (\ref{s29}). The measurement results also match well with the far-field beamforming path loss model given in (\ref{s28}), since the mmWave RIS1 is configured to realize this operation. In addition, we observe that the path loss model given in (\ref{s2}) overestimates the received power. The gap between (\ref{s2}) and (\ref{s29}) is mainly determined by the fact that the size of a single unit cell of the mmWave RIS1 is ${d_x}{d_y} = \frac{\lambda }{8} \times \frac{\lambda }{4}$, which corresponds to a deep sub-wavelength structure. In this case, replacing the factor $\frac{{G{\lambda ^2}}}{{4\pi }} = \frac{{{\lambda ^2}}}{\pi }$ with ${d_x}{d_y} = \frac{{{\lambda ^2}}}{{32}}$ yields a better estimate result.
In addition, by comparing Fig. \ref{mmWaveRIS1mea1}(a) with Fig. \ref{mmWaveRIS1mea1}(c) and Fig. \ref{mmWaveRIS1mea1}(b) with Fig. \ref{mmWaveRIS1mea1}(d), we observe that the received power is angle-dependent, i.e., the received power decreases as $\theta_{t}$ and/or $\theta_{r}$ increase.
\subsubsection{Intelligent Reflection via the mmWave RIS1}\label{IntelligentmmWaveRIS1}
\begin{figure}
\centering
\includegraphics[scale = 0.525]{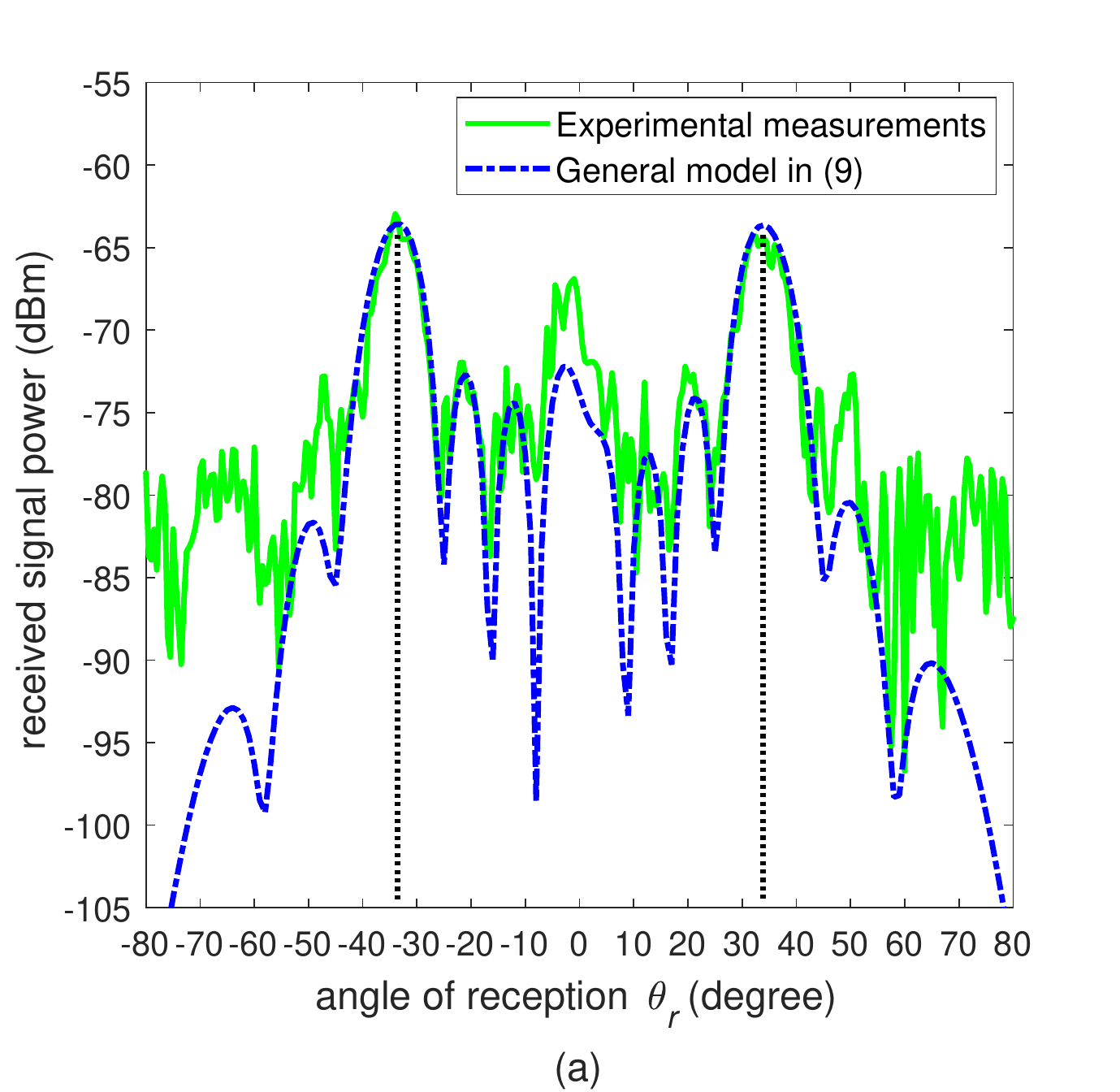}\hspace{-0.5cm}
\includegraphics[scale = 0.525]{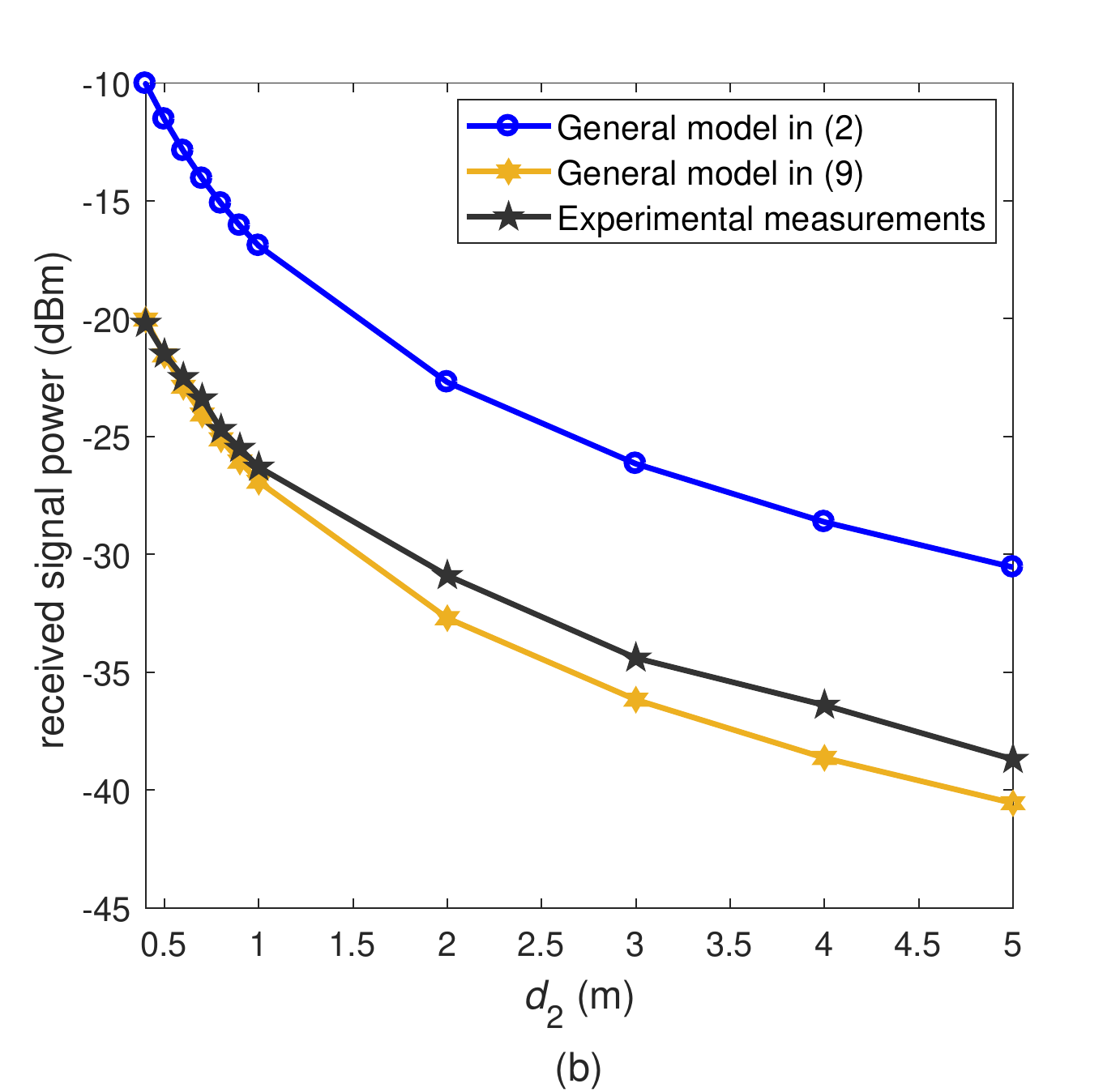}
\caption{Intelligent reflection via mmWave RIS1: Measurements vs. modeling. (a) Received signal power versus $\theta_{r}$ when $\theta_{t}=0$, $d_{1}=1.3\ $m, and $d_{2}=2.6\ $m. (b) Received signal power versus $d_2$ when $\theta_{t}=0$, $\theta_{r}=34^\circ$ and $d_1=0.5\ $m.}
\label{mmWaveRIS1mea2}
\end{figure}
In this case, the experiment is conducted by using the measurement system A. The measurement setup is $P_t=20\ $dBm, $f=27\ $GHz, $\theta_{t}=0$, $d_{1}=1.3\ $m, and $d_{2}=2.6\ $m. The coding state of the unit cells of the mmWave RIS1 is chosen as follows: the coding state ``0'' is applied to $\rm{U}_{n,m}$ if $\bmod (m,14) \in \left[ {1,7} \right]$, and the coding state ``1'' is applied to $\rm{U}_{n,m}$ if $\bmod (m,14) \notin \left[ {1,7} \right]$. This binary coding structure results in a stripe coding pattern of the unit cells and yields an RIS that performs dual-beam intelligent reflection. Further information on the used stripe coding pattern to realize dual-beam reflection is available in \cite{MetaCoding} and \cite{MetaInfo}. Among the many possible options to realize digitally-coded RISs, stripe coding is just an example that we use to validate the accuracy of the general path loss model in (\ref{s29}). Fig. \ref{mmWaveRIS1mea2}(a) illustrates the measured power distribution of the received signal as a function of $\theta_{r}$. By using the proposed coding pattern, we observe that the dual-beam intelligent reflection is successfully realized, and that, in particular, the direction of the two beams is $\left| {{\theta _r}} \right| = {34^\circ }$. The experimental measurements reported in Fig. \ref{mmWaveRIS1mea2}(a) are in good agreement with the proposed general path loss model given in (\ref{s29}). The received signal power as a function of $d_2$ is illustrated in Fig. \ref{mmWaveRIS1mea2}(b). In this case, the measurement system B is employed by setting $\theta_{t}=0$, $\theta_{r}=34^\circ$, and $d_1=0.5\ $m. Once again, we observe that the measurements are in good agreement with the path loss model in (\ref{s29}), which yields more accurate estimates than (\ref{s2}).
\subsection{Validation via Measurement Results by Using the mmWave RIS2}\label{MeasurementmmWaveRIS2}
In this subsection, we consider the mmWave RIS2 and test the accuracy of the proposed path loss model when the metasurface is configured for specular reflection and intelligent reflection.
\subsubsection{Specular Reflection via the mmWave RIS2}\label{SpecularmmWaveRIS2}
\begin{figure}
\centering
\includegraphics[scale = 0.55]{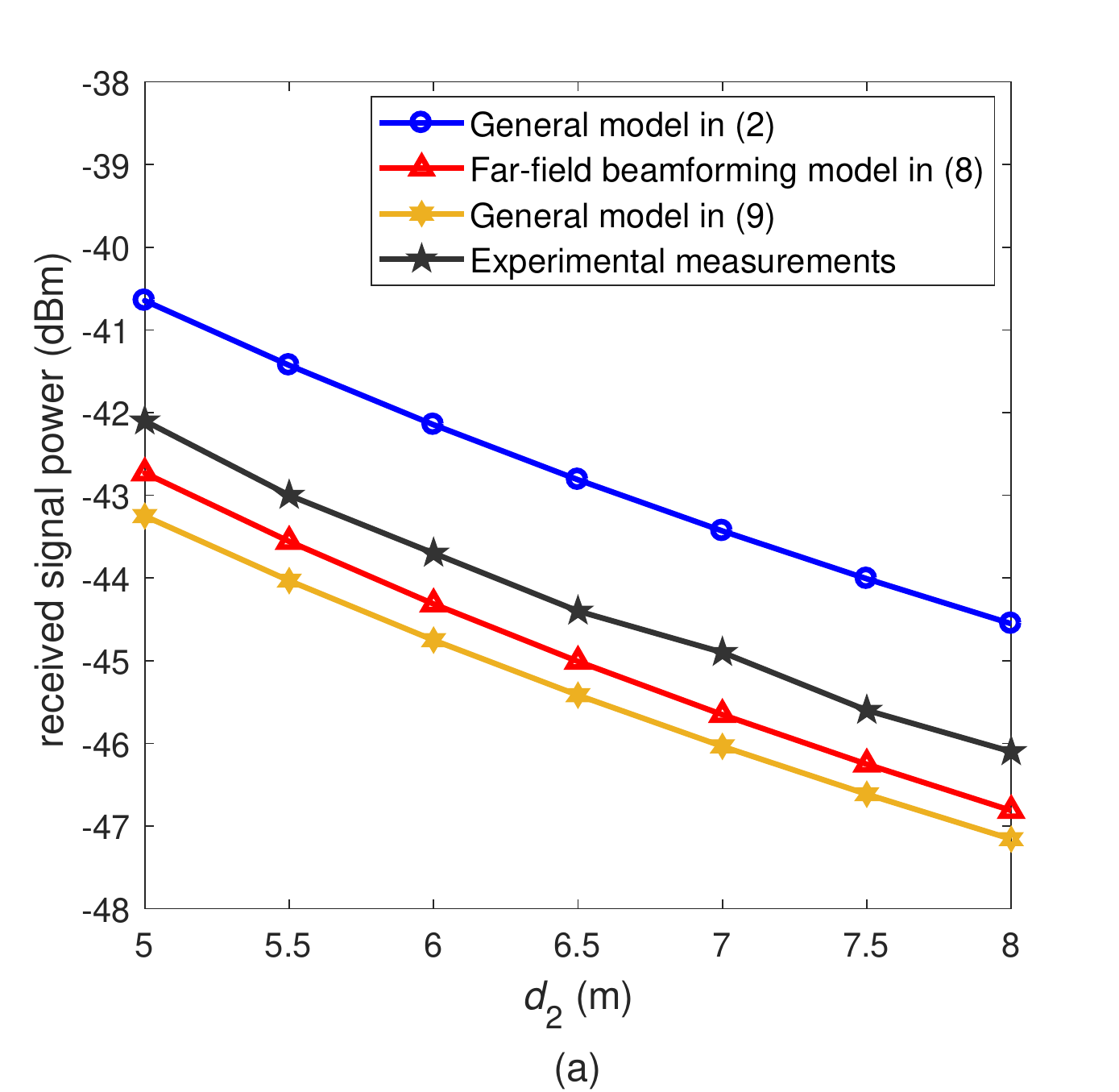}\hspace{-0.5cm}
\includegraphics[scale = 0.55]{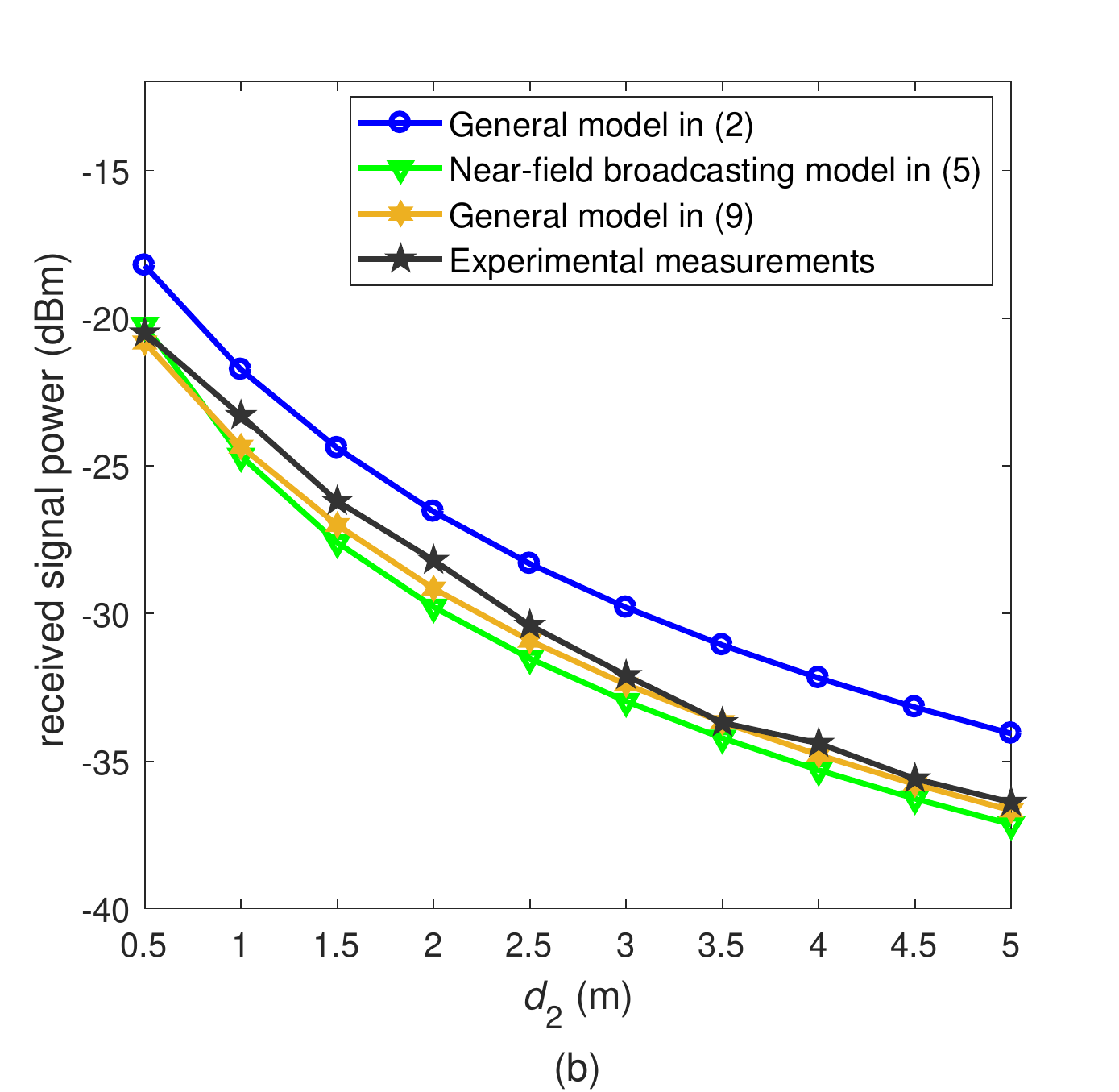}
\caption{Specular reflection via mmWave RIS2: Measurements vs. modeling. (a) Received signal power versus $d_2$ when $d_1=5\ $m and $\theta_{t}=\theta_{r}=10^\circ$. (b) Received signal power versus $d_2$ when $d_1=0.25\ $m and $\theta_{t}=\theta_{r}=10^\circ$.}
\label{mmWaveRIS2mea1}
\end{figure}

The measurement setup is $P_t=20\ $dBm, $f=33\ $GHz, $\theta_{t}=\theta_{r}=10^\circ$, and all the unit cells of mmWave RIS2 are configured in the coding state ``0''. By letting $d_{1}=5\ $m and $d_1=0.25\ $m, we measure the received power as a function of $d_{1}$ and $d_{2}$ by using the measurement system B. Fig. \ref{mmWaveRIS2mea1}(a) illustrates the measured received signal power, as a function of $d_2$, when $d_{1}=5\ $m and $d_{2}\ge5\ $m. Based on Table \ref{parasummary}, in this case, the mmWave RIS2 operates in the far-field region and is configured to realize far-field specular beamforming. We observe that the experimental measurements are in good agreement with the path loss models given in (\ref{s29}) and in (\ref{s28}) for far-field beamforming. Fig. \ref{mmWaveRIS2mea1}(b) reports the measured received power as a function of $d_2$ when $d_{1}=0.25\ $m. In this case, according to Table \ref{parasummary}, the mmWave RIS2 operates in the near-field region and realizes near-field specular broadcasting. The experimental measurements confirm that path loss models in (\ref{s29}) and (\ref{s5}) for near-field broadcasting yield good estimates of the path loss in the considered setup. Unlike the experimental results obtained with the mmWave RIS1, we observe that the path loss model in (\ref{s2}), originally reported in \cite{pathloss}, is sufficiently accurate for the mmWave RIS2. In fact, the absolute error difference is less than 2 dB in the considered setup. In contrast to Fig. \ref{mmWaveRIS1mea1}, the relatively small gap between (\ref{s2}) and (\ref{s29}) in Fig. \ref{mmWaveRIS2mea1} is determined by the fact that the size of a single unit cell of the mmWave RIS2 is approximately equal to half of the wavelength. From Table \ref{parasummary}, more precisely, we have ${d_x}{d_y} = \frac{{{\lambda ^2}}}{{5.7}}$. This factor is not very different from $\frac{{G{\lambda ^2}}}{{4\pi }} = \frac{{{\lambda ^2}}}{\pi }$ that is used in (\ref{s2}) according to \cite{pathloss}. This confirms that the general path loss model in (\ref{s29}) is more accurate for deep sub-wavelength RIS structures.
\subsubsection{Intelligent Reflection via the mmWave RIS2}\label{IntelligentmmWaveRIS2}
\begin{figure}
\centering
\includegraphics[scale = 0.55]{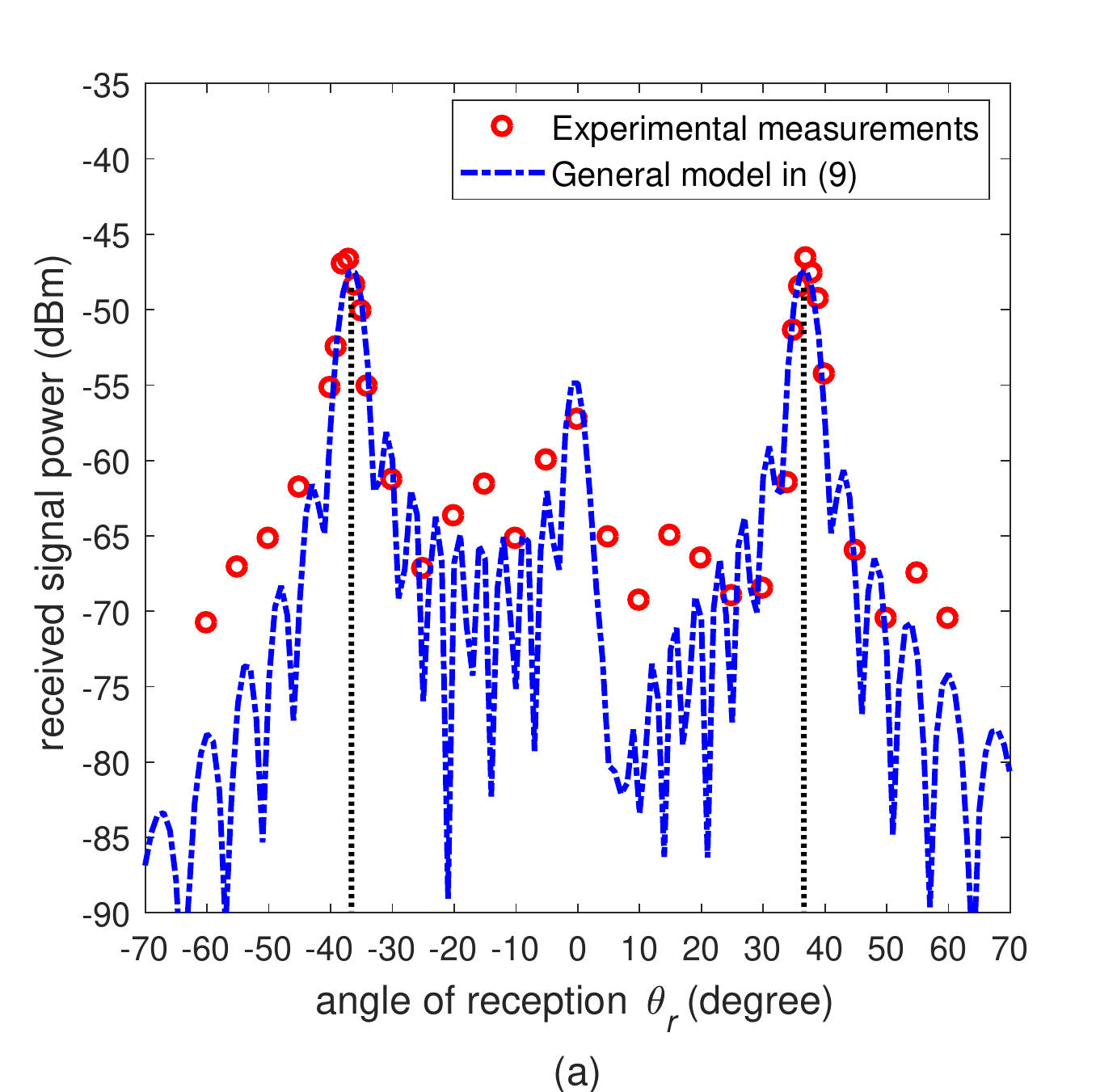}\hspace{-0.5cm}
\includegraphics[scale = 0.55]{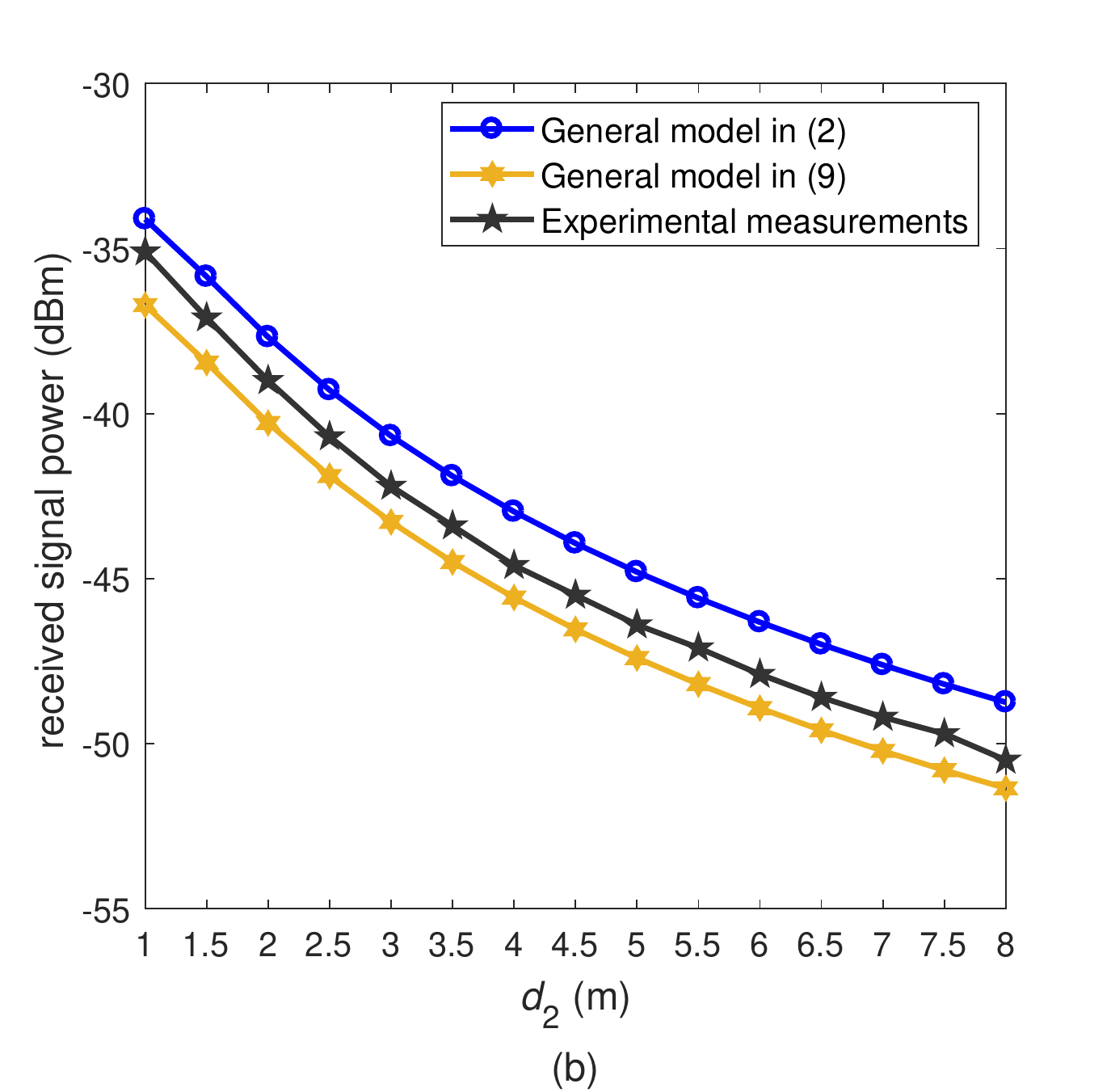}
\caption{Intelligent reflection via mmWave RIS2: Measurements vs. modeling. (a) Received signal power versus $\theta_{r}$ when $\theta_{t}=0$, $d_{1}=5\ $m, and $d_{2}=5\ $m. (b) Received signal power versus $d_2$ when $\theta_{t}=0$, $\theta_{r}=37^\circ$ and $d_1=5\ $m.}
\label{mmWaveRIS2mea2}
\end{figure}

The measurement setup is $P_t=20\ $dBm, $f=33\ $GHz, $\theta_{t}=0$, $d_{1}=5\ $m, and $d_{2}=5\ $m. To realize dual-beam intelligent reflection similar to the mmWave RIS1, the stripe coding pattern of the mmWave RIS2 is the following: The coding state ``0'' is applied to $\rm{U}_{n,m}$ if $\bmod (m,4) \in \left[ {1,2} \right]$, and the coding state ``1'' is applied to $\rm{U}_{n,m}$ if $\bmod (m,4) \notin \left[ {1,2} \right]$. Fig. \ref{mmWaveRIS2mea2}(a) illustrates the measured received signal power as a function of $\theta_{r}$. We observe that the dual-beam intelligent reflection is successfully implemented, and the direction of the two beams is $\left| {{\theta _r}} \right| = {37^\circ }$. From Fig. \ref{mmWaveRIS2mea2}(a), in addition, we evince that the experimental measurements are in good agreement with the general path loss model given in (\ref{s29}). Fig. \ref{mmWaveRIS2mea2}(b) reports the measured received signal power as a function of $d_2$ when $\theta_{t}=0$, $\theta_{r}=37^\circ$, and $d_1=5\ $m. The measurements match well with the general path loss models given in (\ref{s29}) and (\ref{s2}).
\subsection{Validation via the Measurements Reported in \cite{pathloss}}\label{MeasurementCite}
In order to further assess the general path loss model in (\ref{s29}), we study its accuracy by using the measurement results reported in \cite{pathloss}. In particular, the three utilized RISs are referred to as large RIS1, large RIS2, and small RIS. Interested readers are referred to \cite{pathloss} for further information on the specifications of these RISs.
\subsubsection{Specular Reflection via the Large RIS1}\label{PreviousLargeRIS1}
\begin{figure}
\centering
\includegraphics[scale = 0.55]{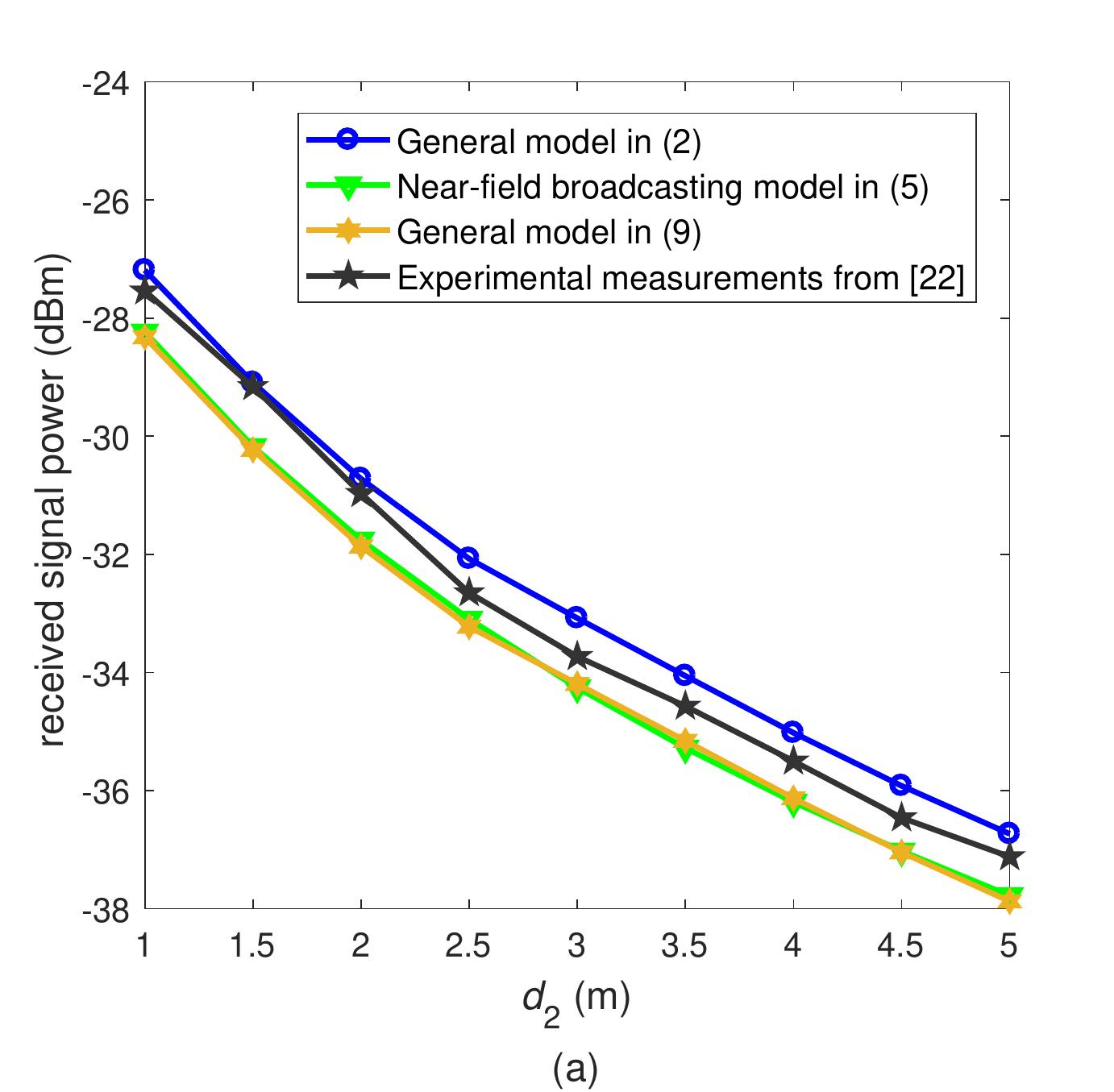}\hspace{-0.5cm}
\includegraphics[scale = 0.55]{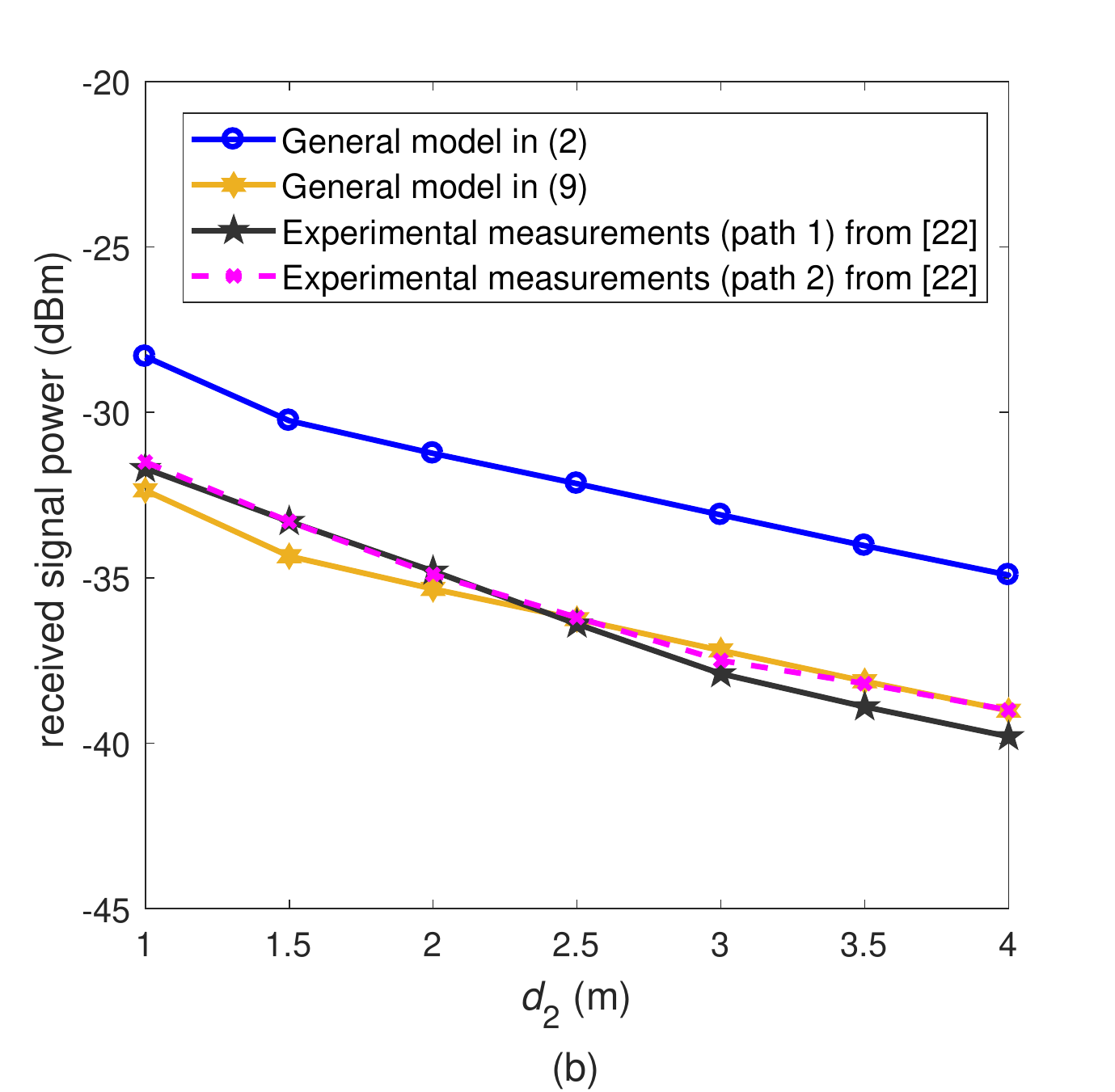}
\caption{Specular and intelligent reflection with large RIS1 and large RIS2: Measurements vs. modeling. (a) Received signal power versus $d_2$ when the large RIS1 performs near-field specular broadcasting, and $d_1=1\ $m, $\theta_{t}=\theta_{r}=45^\circ$. (b) Received signal power versus $d_2$ when the large RIS2 performs intelligent reflection, and $d_1=1\ $m, $\theta_{t}=0$, $\theta_{r}=45^\circ$.}
\label{previousmea1and2}
\end{figure}

The large RIS1 operates at 10.5 GHz and it is configured to realize near-field specular broadcasting. Fig. \ref{previousmea1and2}(a) reproduces the experimental measurements of the received signal power obtained in \cite{pathloss} as a function of $d_2$ when $d_1=1\ $m and $\theta_{t}=\theta_{r}=45^\circ$. We observe that the measurement results are in good agreement with the path loss model in (\ref{s29}) as well.
\begin{figure}
\centering
\includegraphics[scale = 0.55]{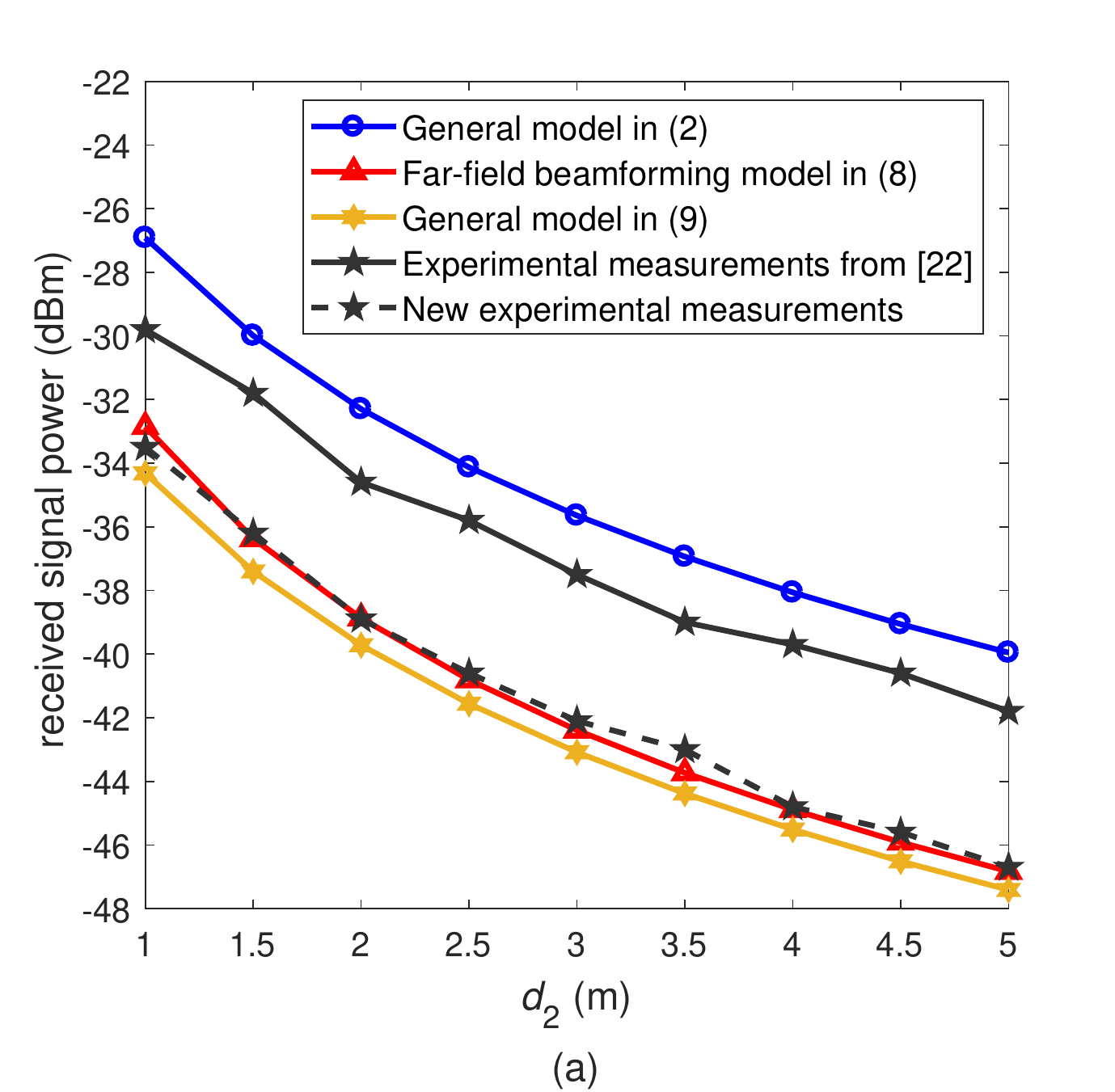}
\includegraphics[scale = 0.37]{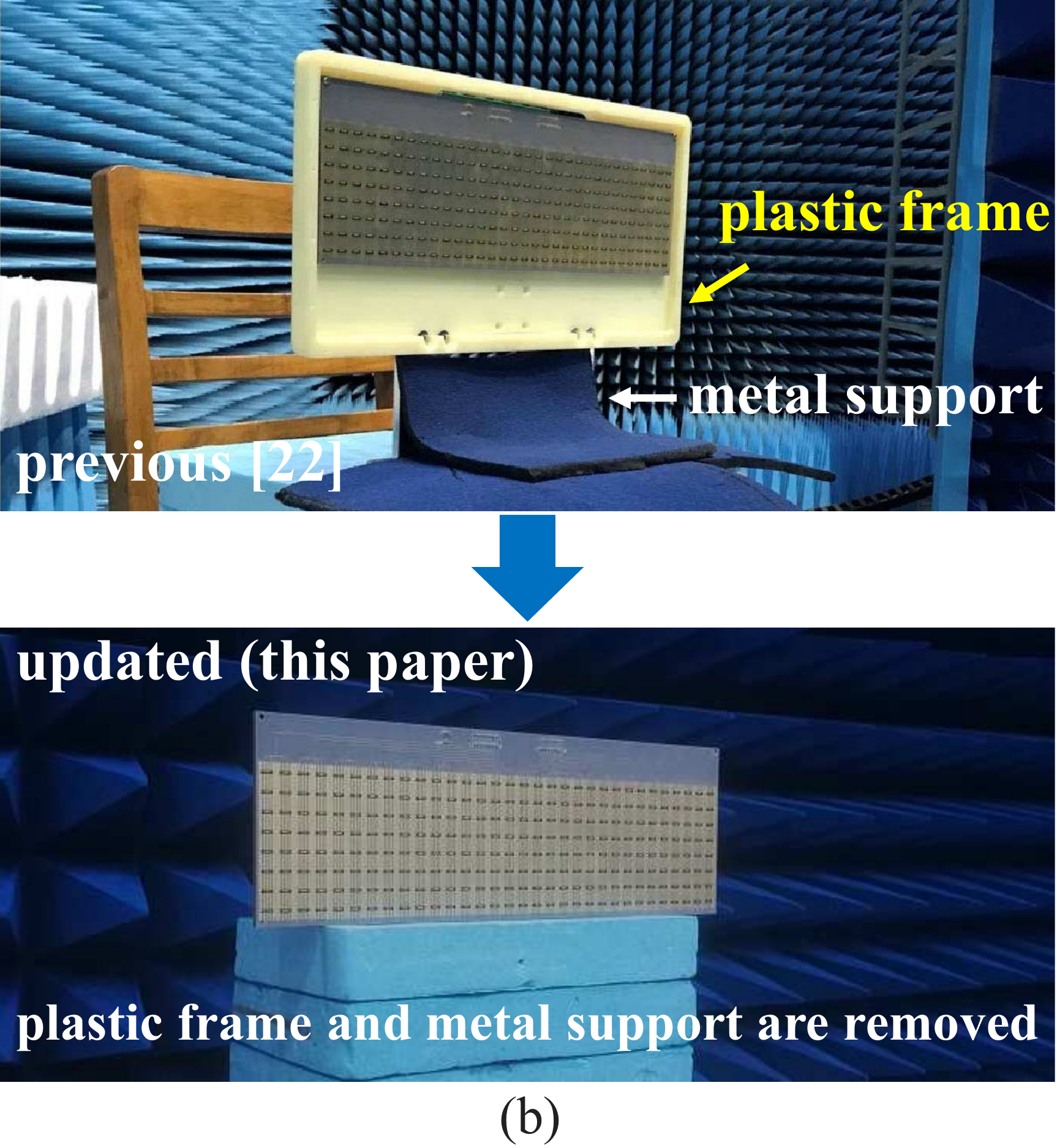}
\caption{Experimental measurements for the small RIS. (a) Received signal power versus $d_2$ when the small RIS performs far-field specular beamforming, and $d_1=1\ $m, $\theta_{t}=\theta_{r}=45^\circ$. (b) The measurement setups in \cite{pathloss} and in this paper.}
\label{previousmea3}
\end{figure}
\subsubsection{Intelligent Reflection via the Large RIS2}\label{PreviousLargeRIS2}
The large RIS2 operates at 10.5 GHz and it is configured to realize intelligent reflection. Fig. \ref{previousmea1and2}(b) reproduces the experimental measurements of the received signal power obtained in \cite{pathloss} as a function of $d_2$ when $d_1=1\ $m, $\theta_{t}=0$, and $\theta_{r}=45^\circ$. We observe that the measurement results are in good agreement with the path loss model in (\ref{s29}). In addition, the path loss model in (\ref{s29}) is more accurate than the path loss model in (\ref{s2}). The relatively good predictions obtained by using (\ref{s2}) can be justified by the fact that the size of the unit cells of the large RIS1 and large RIS2 is approximately half of the wavelength at the operating frequency of 10.5 GHz.

\subsubsection{Specular Reflection via the Small RIS}\label{PreviousSmallRIS}
The small RIS described in \cite{pathloss} is a metasurface structure in which the unit cells have a deep sub-wavelength size. In particular, the size of each unit cell is approximately $\frac{\lambda }{6} \times \frac{\lambda }{6}$ at the operating frequency of 4.25 GHz. The small RIS is configured to perform far-field specular beamforming. Fig. \ref{previousmea3}(a) reproduces the experimental measurements of the received signal power obtained in \cite{pathloss} as a function of $d_2$ when $d_1=1\ $m and $\theta_{t}=\theta_{r}=45^\circ$. In this case, we observe that the path loss model in (\ref{s29}) would be less accurate than the path loss model in (\ref{s2}). After careful inspection, we have identified that the cause of the gap between the previous measurements and the path loss model in (\ref{s29}) is the plastic frame and the metal support that were used in the measurement campaign in \cite{pathloss}. In order to support this intuition, we have conducted new experimental measurements by removing the plastic frame and the metal support. The corresponding results are illustrated in Fig. \ref{previousmea3}(a), which confirms the good accuracy provided by the path loss model in (\ref{s29}) and the better fit that it yields as compared with the path loss model in (\ref{s2}). Once again, the inaccuracy of (\ref{s2}) is related to the deep sub-wavelength structure of the unit cells that constitute the small RIS.

Based on all the measurement results reported in this section, we conclude that the path loss model in (\ref{s29}) yields sufficiently good estimates of the power scattered by an RIS, as a function of several key design parameters and configurations of the unit cells.
\section{Conclusion}\label{Conclusion}
In this paper, we have introduced a simple yet sufficiently accurate free-space path loss model for analyzing RIS-assisted wireless communications. The proposed path loss model has been validated through experimental measurements conducted in the mmWave frequency band by using metasurface-based RISs. Furthermore, the experimental results have validated the relation between the scattering gain of a unit cell and its size, and have suggested an effective expression to model the power radiation pattern of a unit cell. The proposed SPA triangle unveils fundamental tradeoffs for deploying RISs in high frequency bands. These results may help researchers to better assess the performance limits of RISs in wireless networks.

\begin{appendices}
\section*{Appendix A - Proof of Proposition 1}\label{A}
By assuming a transmit antenna with 100$\%$ power efficiency, the antenna gain can be formulated as follows\cite{Book1}
\begin{equation}\label{s7}
G_t = \frac{{4\pi }}{{\int\limits_{\varphi  = 0}^{2\pi } {\int\limits_{\theta  = 0}^\pi  {{{ {F^{tx}\left( {\theta ,\varphi } \right)} }}\sin \theta d\theta d\varphi } } }}.
\end{equation}
A sufficiently general model for the normalized power radiation pattern of a transmit antenna is given by the following equation\cite{Book1}
\begin{equation}\label{s9}
F^{tx}\left( {\theta ,\varphi } \right) = \left\{
\begin{array}{rcl}
{\left(\cos\theta\right)}^\alpha & & {\theta  \in \left[ {0,\frac{\pi }{2}} \right], \varphi  \in \left[ {0,2\pi } \right]}\\
0 & & {\theta  \in \left( {\frac{\pi }{2},\pi } \right], \varphi  \in \left[ {0,2\pi } \right]}
\end{array} \right.
\end{equation}
By inserting (\ref{s9}) in (\ref{s7}), we obtain
\begin{equation}\label{s10}
{G_t} = \frac{{4\pi }}{{\int\limits_{\varphi  = 0}^{2\pi } {\int\limits_{\theta  = 0}^\pi  {{F^{tx}}\left( {\theta ,\varphi } \right)\sin \theta d\theta d\varphi } } }} = \frac{{4\pi }}{{\int\limits_{\varphi  = 0}^{2\pi } {d\varphi \int\limits_{\theta  = 0}^{0.5\pi } {{{\left(\cos\theta\right)}^\alpha} \sin \theta d\theta } } }} = 2\left( {\alpha  + 1} \right).
\end{equation}
Therefore, the normalized power radiation pattern of the transmit antenna can be explicitly formulated in terms of the antenna gain $G_t$ by inserting (\ref{s10}) in (\ref{s9}), which yields
\begin{equation}\label{s11}
F^{tx}\left( {\theta ,\varphi } \right) = \left\{
\begin{array}{rcl}
{\left(\cos\theta\right)}^{(\frac{G_t}{2} - 1)} & & {\theta  \in \left[ {0,\frac{\pi }{2}} \right], \varphi  \in \left[ {0,2\pi } \right]}\\
0 & & {\theta  \in \left( {\frac{\pi }{2},\pi } \right], \varphi  \in \left[ {0,2\pi } \right]}
\end{array} \right.
\end{equation}
Similarly, the normalized power radiation pattern of the receive antenna can be written as follows
\begin{equation}\label{s12}
F^{rx}\left( {\theta ,\varphi } \right) = \left\{
\begin{array}{rcl}
{\left(\cos\theta\right)}^{(\frac{G_r}{2} - 1)} & & {\theta  \in \left[ {0,\frac{\pi }{2}} \right], \varphi  \in \left[ {0,2\pi } \right]}\\
0 & & {\theta  \in \left( {\frac{\pi }{2},\pi } \right], \varphi  \in \left[ {0,2\pi } \right]}
\end{array} \right.
\end{equation}
It is worth noting that the power radiation patterns given in (\ref{s11}) and (\ref{s12}) assume that the transmit/receive antennas radiate/sense the signals to/from only half of the space. Commonly used antennas, such as patch and horn antennas, usually fulfill this assumption.
If the transmit and receive antennas are isotropic, then (\ref{s11}) and (\ref{s12}) need to be recomputed as $F^{tx}\left( {\theta ,\varphi } \right){=}F^{rx}\left( {\theta ,\varphi } \right){=}1$ for ${\theta  \in \left[ {0,\pi} \right], \varphi  \in \left[ {0,2\pi } \right]}$, and $G_t{=}G_r{=}1$ holds true. In addition, (\ref{s11}) and (\ref{s12}) can be replaced by other (even experimental) antenna radiation patterns that are widely used in wireless communications.

As shown in Fig. \ref{systemdescription}, by applying the law of cosines, we have
\begin{equation}\label{s16}
\cos \theta _{n,m}^{tx} = \frac{{{{\left( {{d_1}} \right)}^2} + {{\left( {r_{n,m}^t} \right)}^2} - {{\left( {{d_{n,m}}} \right)}^2}}}{{2{d_1}r_{n,m}^t}},
\end{equation}
and
\begin{equation}\label{s19}
\cos \theta _{n,m}^{rx} = \frac{{{{\left( {{d_2}} \right)}^2} + {{\left( {r_{n,m}^r} \right)}^2} - {{\left( {{d_{n,m}}} \right)}^2}}}{{2{d_2}r_{n,m}^r}}.
\end{equation}

In this paper, we assume the following normalized power radiation pattern for a single unit cell of the RIS
\begin{equation}\label{s20}
F\left( {\theta ,\varphi } \right) = \left\{
\begin{array}{rcl}
{\cos\theta} & & {\theta  \in \left[ {0,\frac{\pi }{2}} \right], \varphi  \in \left[ {0,2\pi } \right]}\\
0 & & {\theta  \in \left( {\frac{\pi }{2},\pi } \right], \varphi  \in \left[ {0,2\pi } \right]}
\end{array} \right.
\end{equation}
In \cite{pathloss}, based on simulation results obtained from the electromagnetic field simulation software, we used the function ${\cos ^3}\theta$ to match the power radiation pattern of the unit cells of three RISs that were employed in the measurement campaign. In general, the radiation pattern of the unit cell is actually related to its specific design and manufacturing. The experimental measurements reported in Section \ref{MeasurementandDiscussion} of this paper suggest differently from \cite{pathloss} that the analytical formulation in (\ref{s20}) is an effective and practical choice for modeling the path loss of RIS-assisted communications.

According to Fig. \ref{systemdescription}, in addition, we have
\begin{equation}\label{s21}
\cos \theta _{n,m}^t = \frac{{{z_t}}}{{r_{n,m}^t}},
\end{equation}
and
\begin{equation}\label{s22}
\cos \theta _{n,m}^r = \frac{{{z_r}}}{{r_{n,m}^r}}.
\end{equation}
Proposition 1 is obtained by substituting (\ref{s11}), (\ref{s12}), (\ref{s16}), (\ref{s19}), (\ref{s20}), (\ref{s21}), and (\ref{s22}) in (\ref{s6}).
\section*{Appendix B - Proof of Proposition 2}\label{B}
Let us consider an RIS that operates in the far-field region and with unit power reflection efficiency, i.e., $\left| {{\Gamma _{n,m}}} \right| = A = 1$. In addition, assume that the angles of incidence and reflection are $\theta_t=\theta_r=0$. Under these assumptions, the RIS acts as a rectangular perfect electric conducting plate, and the received power in (\ref{s1}) simplifies as follows
\begin{equation}\label{s24}
{P_r} = {P_t}\frac{{{G_t}{G_r}G{M^2}{N^2}{d_x}{d_y}{\lambda ^2}}}{{64{\pi ^3}{{({d_1}{d_2})}^2}}}.
\end{equation}
Under the same assumptions, the radar equation can be formulated as follows\cite{Book2}
\begin{equation}\label{s25}
{P_r} = {P_t}\frac{{{G_t}{G_r}{\lambda ^2}\sigma }}{{64{\pi ^3}{{({d_1}{d_2})}^2}}},
\end{equation}
where $\sigma$ denotes the radar cross section (RCS). The RCS of a rectangular metal plate of the same size as the RIS shown in Fig. \ref{systemdescription} can be expressed as\cite{Book2}
\begin{equation}\label{s26}
\sigma  = \frac{{4\pi {{\left( {MN{d_x}{d_y}} \right)}^2}}}{{{\lambda ^2}}}.
\end{equation}
By substituting (\ref{s26}) in (\ref{s25}), we obtain
\begin{equation}\label{s27}
{P_r} = {P_t}\frac{{{G_t}{G_r}{{\left( {MN{d_x}{d_y}} \right)}^2}}}{{16{\pi ^2}{{({d_1}{d_2})}^2}}}
\end{equation}

By comparing (\ref{s24}) and (\ref{s27}), we obtain $G = \frac{{4\pi {d_x}{d_y}}}{{{\lambda ^2}}}$, which yields an explicit relation between the scattering gain and the surface area of a single unit cell of the RIS. Proposition 2 follows from (\ref{s1}) by replacing $G$ with the obtained expression as a function of the surface area of the unit cell, for generic angles of incidence and reflection, in the far-field beamforming scenario.
\end{appendices}

%

\end{document}